\documentclass[journal=jctcce,manuscript=article,layout=twocolumn]{achemso}

\usepackage[colorlinks,linkcolor=blue,citecolor=blue,urlcolor=black,bookmarks=false,hypertexnames=true]{hyperref} 
\usepackage[version=3]{mhchem}
\usepackage{amssymb}
\usepackage{color}
\usepackage{braket}
\usepackage{xspace}
\usepackage{cleveref}
\usepackage{graphicx}
\usepackage{subfig}
\usepackage{threeparttable}
\usepackage{textcomp}

\usepackage{setspace}

\usepackage{array}
\newcolumntype{L}[1]{>{\raggedright\let\newline\\\arraybackslash\hspace{0pt}}m{#1}}
\newcolumntype{C}[1]{>{\centering\let\newline\\\arraybackslash\hspace{0pt}}m{#1}}
\newcolumntype{R}[1]{>{\raggedleft\let\newline\\\arraybackslash\hspace{0pt}}m{#1}}

\makeatletter
\let\l@addto@macro\relax
\makeatother
\usepackage[fontsize=11pt]{scrextend}

\let\oldmaketitle\maketitle
\let\maketitle\relax
\newcommand{\h}[2]{h_{{#1}}^{{#2}}}
\newcommand{\f}[2]{f_{{#1}}^{{#2}}}

\renewcommand{\v}[2]{{v}_{{#1}}^{{#2}}}

\renewcommand{\c}[1]{a^\dagger_{#1}}
\renewcommand{\a}[1]{a_{#1}}

\newcommand{\ta}[1]{\tilde{a}_{#1}}
\newcommand{\pdm}[2]{\gamma_{{#1}}^{{#2}}}

\newcommand{\e}[1]{\ensuremath{\varepsilon_{#1}}}

\newcommand*{\degree}{\ensuremath{^\circ}\xspace}

\newcommand{\angstrom}{\mbox{\normalfont\AA}\xspace}
\newcommand*{\maxe}{$\Delta_{\mathrm{MAX}}$\xspace}
\newcommand*{\mae}{\ensuremath{\Delta_{\mathrm{MAE}}}\xspace}
\newcommand*{\std}{\ensuremath{\Delta_{\mathrm{STD}}}\xspace}
\newcommand*{\teth}{$t$-\ce{C2H4}\xspace}
\newcommand*{\tethp}{$t$-\ce{C2H4}$^+$\xspace}
\newcommand*{\tethm}{$t$-\ce{C2H4}$^-$\xspace}

\crefname{figure}{Figure}{Figures}
\crefname{table}{Table}{Tables}
\crefname{equation}{Eq.}{Eqs.}
\crefname{section}{Section}{Sections}
\crefname{subsection}{Section}{Sections}

\author{Koushik Chatterjee}
\affiliation{%
     Department of Chemistry and Biochemistry,
     The Ohio State University,
     Columbus, Ohio 43210, United States
}
 \author{Alexander Yu.\ Sokolov}
 \email{sokolov.8@osu.edu}
 \affiliation{%
     Department of Chemistry and Biochemistry,
     The Ohio State University,
     Columbus, Ohio 43210, United States
 }

\begin{tocentry}
\includegraphics{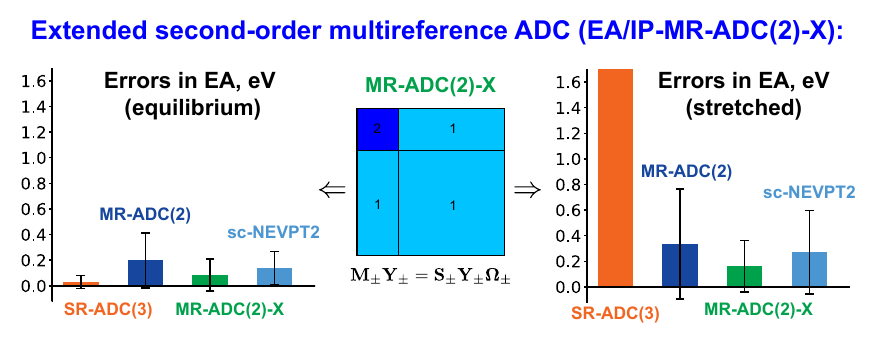}
\end{tocentry}

\title{Extended Second-Order Multireference Algebraic Diagrammatic Construction Theory for Charged Excitations}

\begin{document}

%ArXiv %%%%%%%%%%%%%%
\newcommand*{\abstractext}{We report a new implementation of multireference algebraic diagrammatic construction theory (MR-ADC) for simulations of electron attachment and ionization in strongly correlated molecular systems (EA/IP-MR-ADC). Following our recent work on IP-MR-ADC [{\it J.\@ Chem.\@ Theory Comput.\@} {\bf 2019}, {\it 15}, 5908], we present the first implementation of the second-order MR-ADC method for electron attachment (EA-MR-ADC(2)), as well as two extended second-order approximations (EA- and IP-MR-ADC(2)-X) that incorporate a partial treatment of third-order electron correlation effects. Introducing a small approximation for the second-order amplitudes of the effective Hamiltonian, our implementation of EA- and IP-MR-ADC(2)-X has a low $\mathcal{O}(M^5)$ computational scaling with the basis set size $M$. Additionally, we describe an efficient algorithm for solving the first-order amplitude equations in MR-ADC and partially-contracted second-order N-electron valence perturbation theory (NEVPT2) that completely avoids computation of the four-particle reduced density matrices without introducing any approximations or imaginary-time propagation. For a benchmark set of eight small molecules, carbon dimer, and a twisted ethylene, we demonstrate that EA- and IP-MR-ADC(2)-X achieve accuracy similar to that of strongly-contracted NEVPT2, while having a lower computational scaling with the active space size and providing efficient access to transition properties.
\vspace{0.25cm}
}
%ArXiv %%%%%%%%%%%%%%

%ArXiv %%%%%%%%%%%%%%
\twocolumn[
\begin{@twocolumnfalse}
\oldmaketitle
\vspace{-0.75cm}
\begin{abstract}
\abstractext
\end{abstract}
\end{@twocolumnfalse}
]
%ArXiv %%%%%%%%%%%%%%

\section{Introduction}
\label{sec:intro}

Recently, we proposed a multireference formulation of algebraic diagrammatic construction theory (MR-ADC) for simulations of electronic excitations and spectra in strongly correlated chemical systems.\cite{Sokolov:2018p204113} MR-ADC is a generalization of the single-reference algebraic diagrammatic construction approach\cite{Schirmer:1982p2395,Schirmer:1983p1237,Schirmer:1991p4647,Mertins:1996p2140,Schirmer:2004p11449,Dreuw:2014p82,Banerjee:2019p224112} that aims to obtain excitation energies and transition probabilities from poles and residues of a retarded propagator approximated using multireference perturbation theory (MRPT). Similar to conventional MRPT,\cite{Wolinski:1987p225,Hirao:1992p374,Werner:1996p645,Finley:1998p299,Andersson:1990p5483,Andersson:1992p1218,Angeli:2001p10252,Angeli:2001p297,Angeli:2004p4043,Kurashige:2011p094104,Kurashige:2014p174111,Guo:2016p1583,Sharma:2017p488,Yanai:2017p4829,Sokolov:2017p244102} MR-ADC uses multiconfigurational (complete active-space) wavefunctions to describe static correlation in frontier (active) molecular orbitals of the ground and excited electronic states and perturbatively treats dynamic correlation in the remaining orbitals. 

However, several important differences between MR-ADC and multistate MRPT exist. Rather than constructing perturbed (dynamically correlated) active-space wavefunctions for each electronic state of interest, as it is done in conventional multistate MRPT,\cite{Finley:1998p299,Angeli:2004p4043} in MR-ADC dynamic correlation information is determined for a single (so-called ``parent'' or ``reference'') state and the differential electron correlation in the remaining states is assumed to be simple to describe. This allows MR-ADC to describe many electronic excitations (including those outside of the active space) with a computational cost lower than that of a single multistate MRPT calculation and removes the need for using reference wavefunctions with state-averaged orbitals, which introduce dependence of results on weights used in state-averaging and can be numerically difficult to compute. Importantly, MR-ADC also provides an efficient route to obtaining various transition properties (such as intensities and spectral densities) that are not directly accessible in conventional MRPT calculations. In its formulation, MR-ADC has a close connection to multireference propagator,\cite{Banerjee:1978p389,Yeager:1979p77,Dalgaard:1980p816,Yeager:1984p85,Graham:1991p2884,Yeager:1992p133,Nichols:1998p293,Khrustov:2002p507,HelmichParis:2019p174121} 
linear-response,\cite{Chattopadhyay:2000p7939,Chattopadhyay:2007p1787,Jagau:2012p044116,Samanta:2014p134108,Kohn:2019p041106} 
and equation-of-motion approaches,\cite{Datta:2012p204107,Nooijen:2014p081102,Huntington:2015p194111} 
but, in contrast to many of these methods, is based on a Hermitian eigenvalue problem, which ensures that the computed excitation energies have real values.

In our earlier work we reported an implementation and benchmark of MR-ADC for simulations of ionization processes in multireference systems that incorporates all contributions to transition energies and spectroscopic amplitudes up to second order in perturbation theory (IP-MR-ADC(2)).\cite{Chatterjee:2019p5908} The IP-MR-ADC(2) method was found to provide reliable results for a variety of weakly and strongly correlated systems, however the computed errors in ionization energies were found to be larger than those of the conventional second-order MRPT. In this manuscript, we expand the applicability of the MR-ADC framework by developing its second-order implementation for simulations of electron attachment (EA-MR-ADC(2)). Additionally, we report two extended second-order MR-ADC approximations (EA- and IP-MR-ADC(2)-X) that incorporate third-order electron correlation effects into computation of charged excitation energies and transition properties. We benchmark the new methods against electron affinities and ionization energies obtained from accurate semi-stochastic heat-bath configuration interaction method and compare their performance with conventional single- and multireference methods. 

\section{Theory}
\label{sec:theory}

\subsection{Overview of MR-ADC}
\label{sec:theory:mr_adc_overview}

We start by reviewing the main aspects of the MR-ADC formulation. For more details on the derivation of MR-ADC using the formalism of effective Liouvillean theory,\cite{Mukherjee:1989p257} the reader is referred to Ref. \citenum{Sokolov:2018p204113}. The central object of interest in MR-ADC is a retarded propagator\cite{Fetter2003,Dickhoff2008} $G_{\mu\nu}(\omega)$ expressed in a general form as:
\begin{align}
	\label{eq:g_munu}
	G_{\mu\nu}(\omega)
	& = G_{\mu\nu}^+(\omega) \pm G_{\mu\nu}^-(\omega) \notag\\
	& =  \bra{\Psi}q_\mu(\omega - H + E)^{-1}q^\dag_\nu\ket{\Psi} \notag \\
	&\pm \bra{\Psi}q^\dag_\nu(\omega + H - E)^{-1}q_\mu\ket{\Psi}
\end{align}
Here, $G_{\mu\nu}(\omega)$ describes the response of a many-electron system in an initial state $\ket{\Psi}$ to an external perturbation with frequency $\omega$. The wavefunction $\ket{\Psi}$ is an eigenstate of the electronic Hamiltonian $H$ with energy $E$ and the frequency can be expressed in terms of its real and imaginary parts $\omega \equiv \omega' + i\eta$. The form of $q^\dag_\nu$ and $q_\mu$, referred to as the perturbation and observable operators, determines the nature of a spectroscopic process described by $G_{\mu\nu}(\omega)$. The operators $H$, $q^\dag_\nu$, and $q_\mu$ are usually expressed in their second-quantized form, where the number of creation and annihilation operators in $q^\dag_\nu$ (odd or even) determines the sign ($+$ or $-$) in \cref{eq:g_munu}. The $G_{\mu\nu}^+(\omega)$ and $G_{\mu\nu}^-(\omega)$ terms in \cref{eq:g_munu} are referred to as the forward and backward components of the propagator, respectively. In this work, we will focus on the propagator with $q^\dag_\nu = \c{p}$ and $q_\mu = \a{q}$ that describes electron attachment and ionization processes in photoelectron spectroscopy, also known as the one-particle Green's function.\cite{Goscinski:1980p385,Nooijen:1992p55,Kowalski:2014p094102,Schirmer:1983p1237,Schirmer:1998p4734,Trofimov:2005p144115,Cederbaum:1975p290,VonNiessen:1984p57,Ortiz:2012p123,Banerjee:1978p389,Yeager:1992p133,Nichols:1998p293,Banerjee:2019p224112}

\begin{figure}[t!]
	\includegraphics[width=0.5\textwidth]{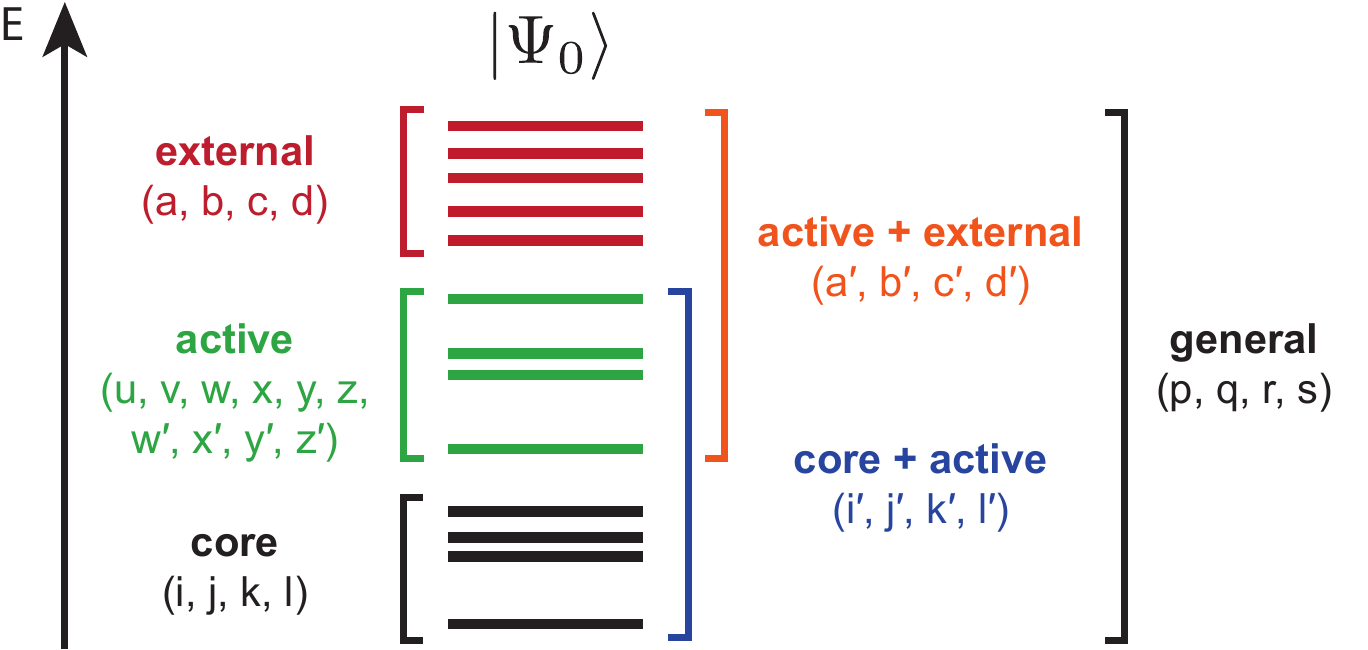}
	\captionsetup{justification=raggedright,singlelinecheck=false}
	\caption{Orbital index convention used in this work.}
	\label{fig:mo_diagram}
\end{figure}

The MR-ADC approach uses multireference perturbation theory (MRPT) to compute accurate approximations to the exact propagator for systems with multiconfigurational nature of the wavefunction. To accomplish this, the molecular orbitals of the system are split into core, active, and external subspaces (\cref{fig:mo_diagram}) and the $N$-electron wavefunction $\ket{\Psi}$ is expressed in terms of the zeroth-order (reference) wavefunction $\ket{\Psi_0}$, obtained from a complete active space configuration interaction (CASCI) or self-consistent field (CASSCF) calculation, as follows:
\begin{align}
	\label{eq:mr_adc_wfn}
	\ket{\Psi} &= e^{A} \ket{\Psi_0} = e^{T - T^\dag} \ket{\Psi_0} , \quad T = \sum_{k=1}^N T_k  \\
	\label{eq:mr_adc_t_amplitudes}
	T_k &= \frac{1}{(k!)^2} {\sum_{i'j'a'b'\ldots}} t_{i'j'\ldots}^{a'b'\ldots} \c{a'}\c{b'}\ldots\a{j'}\a{i'}
\end{align}
In \cref{eq:mr_adc_wfn}, the unitary wave operator $e^{A}$ ($A^\dag = -A$) is parametrized in terms of the amplitudes of the excitation operator $T$ that generates all internally-contracted excitations between core, active, and external orbitals.\cite{Kirtman:1981p798,Hoffmann:1988p993,Yanai:2006p194106,Chen:2012p014108,Li:2015p2097} 

To construct perturbative approximations to $G_{\mu\nu}(\omega)$, the total electronic Hamiltonian $H$ is separated into its zeroth-order $H^{(0)}$ and perturbation $V$ contributions. The $H^{(0)}$ operator is chosen to be the Dyall Hamiltonian\cite{Dyall:1995p4909,Angeli:2001p10252,Angeli:2004p4043}
\begin{align}
	\label{eq:h_dyall_general}
	H^{(0)} &\equiv C + \sum_{i} \e{i} \c{i}\a{i} + \sum_{a} \e{a} \c{a}\a{a} + H_{act} \\
	\label{eq:h_act}
	H_{act} &= \sum_{xy}(\h{x}{y} + \sum_{i} \v{xi}{yi}) \c{x} \a{y}
	+ \frac{1}{4} \sum_{xywz} \v{xy}{zw} \c{x} \c{y} \a{w} \a{z} \\
	C &= \sum_i \h{i}{i} + \frac{1}{2}\sum_{ij}\v{ij}{ij} - \sum_i  \e{i} \\
	\label{eq:f_gen}
	\f{p}{q} &= \h{p}{q} + \sum_{rs} \v{pr}{qs} \pdm{s}{r} \ , \quad \pdm{q}{p} = \braket{\Psi_0|\c{p}\a{q}|\Psi_0}
\end{align}
expressed in the basis of molecular orbitals that diagonalize the core and external blocks of the generalized Fock matrix \eqref{eq:f_gen} with eigenvalues $\e{i}$ and $\e{a}$, respectively. Expanding the propagator $G_{\mu\nu}(\omega)$ in the multireference perturbative series and truncating the expansion at the $n$th order defines the propagator of the MR-ADC(n) approximation:
\begin{align}
	\label{eq:g_pt_series}
	\mathbf{G}(\omega) & \approx \mathbf{G}^{(0)}(\omega) + \mathbf{G}^{(1)}(\omega) + \ldots + \mathbf{G}^{(n)}(\omega)
\end{align}

A special feature of MR-ADC is that the forward and backward components of the propagator ($G_{\mu\nu}^+(\omega)$ and $G_{\mu\nu}^-(\omega)$ in \cref{eq:g_munu}) are decoupled at any level of approximation. As a result, the perturbative expansion \eqref{eq:g_pt_series} can be performed for each component independently. In practice, the forward and backward contributions to the MR-ADC(n) propagator \eqref{eq:g_pt_series} are expressed in the matrix form:
\begin{align}
	\label{eq:Gn_matrix}
	\mathbf{G}_{\pm}(\omega) & = \mathbf{T}_{\pm} \left(\omega \mathbf{S}_{\pm} - \mathbf{M}_{\pm}\right)^{-1} \mathbf{T}_{\pm}^{\dag}
\end{align}
where $\mathbf{M}_{\pm}$, $\mathbf{T}_{\pm}$, and $\mathbf{S}_{\pm}$ are the effective Liouvillean, transition moment, and overlap matrices, respectively, each evaluated up to the $n$th order in perturbation theory. The MR-ADC(n) transition energies are obtained as eigenvalues ($\boldsymbol{\Omega}_{\pm}$) of the Hermitian matrix $\mathbf{M}_{\pm}$ by solving a generalized eigenvalue problem
\begin{align}
	\label{eq:adc_eig_problem}
	\mathbf{M}_{\pm} \mathbf{Y}_{\pm}  = \mathbf{S}_{\pm} \mathbf{Y}_{\pm} \boldsymbol{\Omega}_{\pm}
\end{align}
The resulting eigenvectors $\mathbf{Y}_{\pm}$ are combined with the effective transition moments matrix $\mathbf{T}_{\pm}$ to compute spectroscopic amplitudes
\begin{align}
	\label{eq:spec_amplitudes}
	\mathbf{X}_{\pm} = \mathbf{T}_{\pm} \mathbf{S}_{\pm}^{-1/2} \mathbf{Y}_{\pm}
\end{align}
which provide information about spectral intensities and can be used to evaluate the MR-ADC(n) propagator and spectral function
\begin{align}
	\label{eq:g_mr_adc}
	\mathbf{G}_{\pm}(\omega) &= \mathbf{X}_{\pm} \left(\omega - \boldsymbol{\Omega}_{\pm}\right)^{-1}  \mathbf{X}_{\pm}^\dag \\
	\label{eq:spec_function}
	T(\omega) &= -\frac{1}{\pi} \mathrm{Im} \left[ \mathrm{Tr} \, \mathbf{G}_{\pm}(\omega) \right]
\end{align}

\subsection{MR-ADC(2) for Electron Attachment and Ionization}
\label{sec:theory:mr_adc_2}

We now discuss the second-order MR-ADC approximations for the forward and backward components of the one-particle Green's function ($\mathbf{G}_{+}(\omega)$ and $\mathbf{G}_{-}(\omega)$) that describe electron attachment and ionization processes (EA- and IP-MR-ADC(2)), respectively. For more details about IP-MR-ADC(2), we refer the reader to our previous publication.\cite{Chatterjee:2019p5908}

In EA- and IP-MR-ADC(2), the $\mathbf{M}_\pm$, $\mathbf{T}_\pm$, and $\mathbf{S}_\pm$ matrices in \cref{eq:Gn_matrix} are evaluated up to the second order in MRPT. The $n$th-order contributions to the EA-MR-ADC(2) matrices have the form:\cite{Mukherjee:1989p257,Sokolov:2018p204113,Banerjee:2019p224112} 
\begin{align}
	\label{eq:M+_matrix}
	M_{+\mu\nu}^{(n)} &= \sum_{klm}^{k+l+m=n} \braket{\Psi_0|[h_{+\mu}^{(k)},[\tilde{H}^{(l)},h_{+\nu}^{(m)\dagger}]]_{+}|\Psi_0} \\
	\label{eq:T+_matrix}
	T_{+p\nu}^{(n)} &= \sum_{kl}^{k+l=n} \braket{\Psi_0|[\ta{p}^{(k)},h_{+\nu}^{(l)\dagger}]_{+}|\Psi_0} \\
	\label{eq:S+_matrix}
	S_{+\mu\nu}^{(n)} &= \sum_{kl}^{k+l=n} \braket{\Psi_0|[h_{+\mu}^{(k)},h_{+\nu}^{(l)\dagger}]_{+}|\Psi_0}
\end{align}                                                                                                                                                                            
where $\tilde{H}^{(k)}$ and $\ta{p}^{(k)}$ are the $k$th-order contributions to the effective Hamiltonian $\tilde{H} = e^{-A} H e^{A}$ and observable $\ta{p} = e^{-A} \a{p} e^{A}$ operators, $h_{+\mu}^{(k)\dagger}$ are $k$th-order electron attachment operators that define the $(N+1)$-electron internally-contracted EA-MR-ADC basis states $\ket{\Psi^{(k)}_{+\mu}} = h_{+\mu}^{(k)\dagger}\ket{\Psi_0}$, and $[\ldots]$ and $[\ldots]_+$ denote commutator and anticommutator, respectively. Similarly, the $n$-th order IP-MR-ADC(2) matrix elements are expressed as:
\begin{align}                                                                                                                                                                          
	\label{eq:M-_matrix}
	M_{-\mu\nu}^{(n)} &= \sum_{klm}^{k+l+m=n} \braket{\Psi_0|[h_{-\mu}^{(k)\dagger},[\tilde{H}^{(l)},h_{-\nu}^{(m)}]]_{+}|\Psi_0}  \\
	\label{eq:T-_matrix}
	T_{-p\nu}^{(n)} &= \sum_{kl}^{k+l=n} \braket{\Psi_0|[\ta{p}^{(k)},h_{-\nu}^{(l)}]_{+}|\Psi_0} \\
	\label{eq:S-_matrix}
	S_{-\mu\nu}^{(n)} &= \sum_{kl}^{k+l=n} \braket{\Psi_0|[h_{-\mu}^{(k)\dagger},h_{-\nu}^{(l)}]_{+}|\Psi_0} 
\end{align}                                                                                                                                                                            
where we additionally introduce the $k$th-order ionization operators $h_{-\mu}^{(k)\dagger}$ that define the $(N-1)$-electron basis states $\ket{\Psi^{(k)}_{-\mu}} = h_{-\mu}^{(k)\dagger}\ket{\Psi_0}$ used for solving the IP-MR-ADC equations. 

\begin{figure*}[t!]
	\subfloat[EA]{\label{fig:excitations_ea}\includegraphics[width=0.8\textwidth]{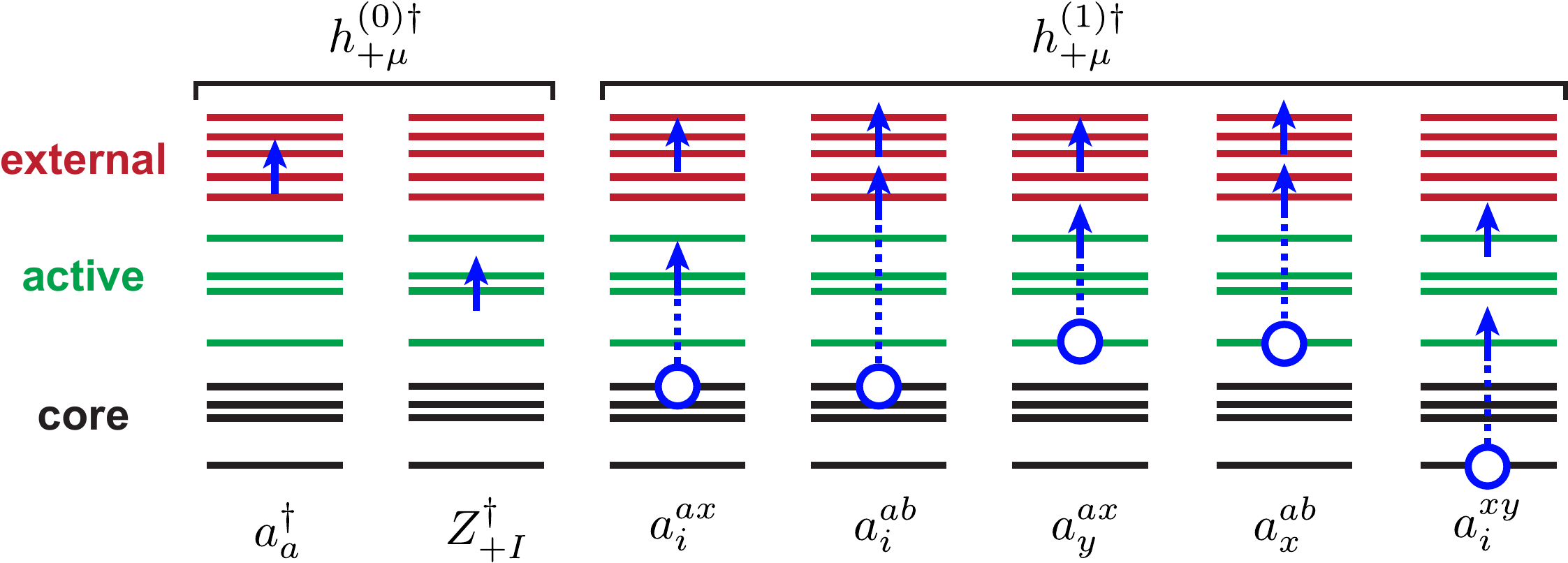}} \qquad \qquad
	\subfloat[IP]{\label{fig:excitations_ip}\includegraphics[width=0.8\textwidth]{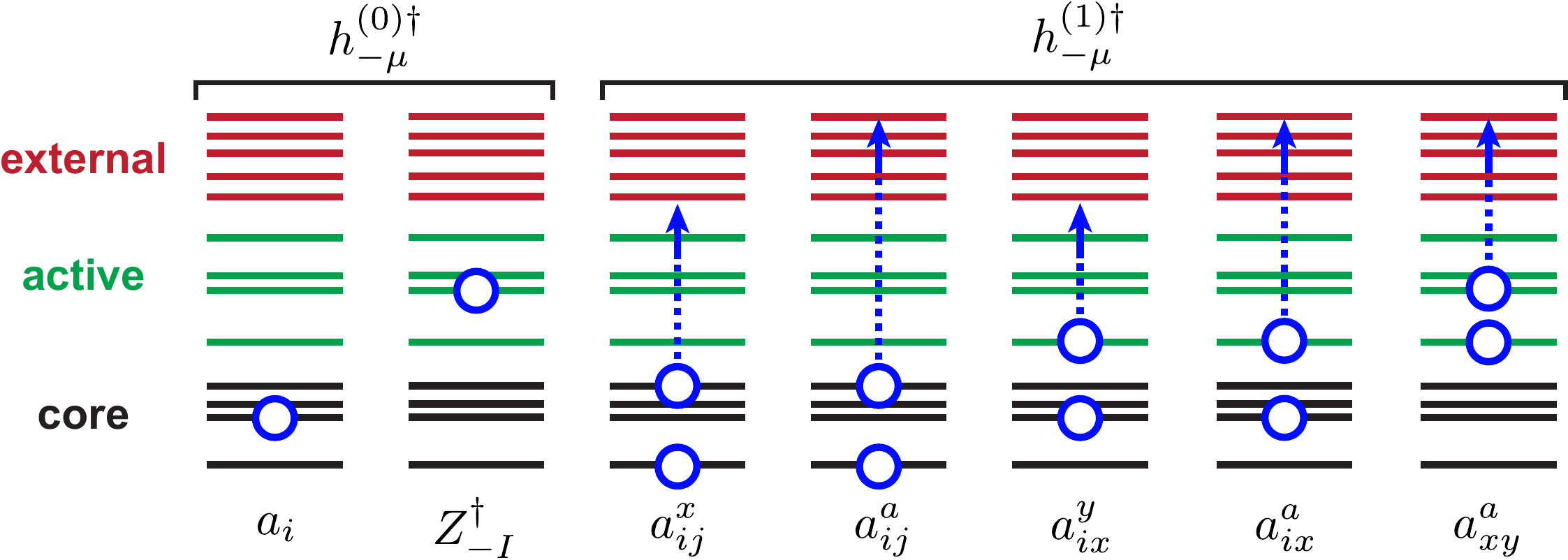}}
	\captionsetup{justification=raggedright,singlelinecheck=false}
	\caption{Schematic illustration of the electron-attached (\ref{fig:excitations_ea}) and ionized (\ref{fig:excitations_ip}) states produced by acting the $h_{\pm\mu}^{(k)\dag}$ ($k = 0, 1$) operators on the reference state $\ket{\Psi_0}$. An arrow represents electron attachment, a circle denotes ionization, and a circle connected with an arrow denotes single excitation. The operators $Z_{\pm I}^\dag$ incorporate all photoelectron transitions in the active orbitals.}
	\label{fig:excitations}
\end{figure*}

\cref{fig:excitations_ea,fig:excitations_ip} illustrate the EA- and IP-MR-ADC(2) basis states obtained by acting the electron attachment $h_{+\mu}^{(k)\dagger}$ and ionization $h_{-\mu}^{(k)\dagger}$ operators on the reference state $\ket{\Psi_0}$. Similar to the single-reference EA- and IP-SR-ADC(2) approximations,\cite{Banerjee:2019p224112} the EA- and IP-MR-ADC(2) equations depend only on the zeroth- and first-order operators $h_{\pm\mu}^{(0)\dagger}$ and $h_{\pm\mu}^{(1)\dagger}$. The zeroth-order operators $h_{\pm\mu}^{(0)\dagger}$ have the form:
\begin{align}
\label{eq:ea_zeroth_order_manifold}
\mathbf{h}_{+}^{(0)\dag} &= \left\{\c{a}; Z_{+I}^\dag \right\}\ , \quad Z_{+I}^\dag = \ket{\Psi_{+I}}\bra{\Psi_0} \\
\label{eq:ip_zeroth_order_manifold}
\mathbf{h}_{-}^{(0)\dag} &= \left\{\a{i}; Z_{-I}^\dag \right\}\ , \quad Z_{-I}^\dag = \ket{\Psi_{-I}}\bra{\Psi_0}
\end{align}
where $h_{+\mu}^{(0)\dagger} = \c{a}$ describe a single-electron attachment in the external orbitals and $Z_{+I}^\dag$ incorporate description of all photoelectron transitions in the active orbitals by projecting the $N$-electron reference state $\ket{\Psi_0}$ onto the $(N+1)$-electron CASCI states $\ket{\Psi_{+I}}$ where the extra electron is added to the active space. Similarly, the operators $h_{-\mu}^{(0)\dagger}$ describe a one-electron ionization in the core orbitals, while $Z_{-I}^\dag$ include ionization and excitations in the active orbitals. Defining $a_{r}^{pq} \equiv \c{p}\c{q}\a{r}$ and $a_{qr}^{p} \equiv \c{p}\a{r}\a{q}$, the first-order operators $h_{\pm\mu}^{(1)\dagger}$ are expressed as:
\begin{align}
\label{eq:ea_first_order_manifold}
\mathbf{h}_{+}^{(1)\dag} &= \left\{a_{i}^{ax}; a_{i}^{ab}; a_{y}^{ax}; a_{x}^{ab}; a_{i}^{xy}\right\} \\
\label{eq:ip_first_order_manifold}
\mathbf{h}_{-}^{(1)\dag} &= \left\{a_{ij}^{x}; a_{ij}^{a}; a_{ix}^{y}; a_{ix}^{a}; a_{xy}^{a}\right\} 
\end{align}
As depicted in \cref{fig:excitations}, these operators describe attachment or ionization of an electron accompanied by a one-electron excitation between core, active, or external orbitals. The $\mathbf{h}_{\pm}^{(1)\dag}$ operators do not contain the all-active operators $a_{z}^{xy}$ and $a_{zy}^{x}$ because all active-space transitions are incorporated by the $Z_{\pm I}^\dag$ operators in $\mathbf{h}_{\pm}^{(0)\dag}$.

Evaluating the EA- and IP-MR-ADC(2) matrix elements in \cref{eq:M+_matrix,eq:T+_matrix,eq:S+_matrix,eq:M-_matrix,eq:T-_matrix,eq:S-_matrix} also requires expressions for the low-order effective operators $\tilde{H}^{(k)}$ and $\ta{p}^{(k)}$ (up to $k$ = 2). These equations are obtained by expanding $\tilde{H}$ and $\ta{p}$ using the Baker--Campbell--Hausdorff (BCH) formula and collecting terms at the $k$th order: 
\begin{align}
	\label{eq:H_bch_0}
	\tilde{H}^{(0)} &= H^{(0)} \\
	\label{eq:H_bch_1}
	\tilde{H}^{(1)} &= V + [H^{(0)}, A^{(1)}]  \\
	\label{eq:H_bch_2}
	\tilde{H}^{(2)} &= [H^{(0)}, A^{(2)}] + \frac{1}{2}[V + \tilde{H}^{(1)}, A^{(1)}] \\
	\label{eq:q_bch_0}
	\ta{p}^{(0)} &= \a{p} \\
	\label{eq:q_bch_1}
	\ta{p}^{(1)} &= [\a{p}, A^{(1)}] \\
	\label{eq:q_bch_2}
	\ta{p}^{(2)} &= [\a{p}, A^{(2)}] + \frac{1}{2} [[\a{p}, A^{(1)}], A^{(1)}]
\end{align}
where $A^{(k)} \equiv T^{(k)} - T^{(k)\dag}$ as defined in \cref{eq:mr_adc_wfn}. \cref{eq:H_bch_0,eq:H_bch_1,eq:H_bch_2,eq:q_bch_0,eq:q_bch_1,eq:q_bch_2} depend on the cluster excitation operators $T^{(k)}$ $(k = 1,2)$ that can be expressed in a general form:
\begin{align}
	\label{eq:excit_op_tensor}
	T^{(k)} = \mathbf{t^{(k)}} \, \boldsymbol{\tau}^\dag =  \sum_\mu t_\mu^{(k)} \tau^\dag_\mu
\end{align}
where the $k$th-order amplitudes $t_\mu^{(k)}$ are contracted with strings of creation and annihilation operators denoted as $\tau^\dag_\mu$  (cf.\@ \cref{eq:mr_adc_t_amplitudes}). The first-order amplitudes $t_\mu^{(1)}$ can be grouped into 11 classes corresponding to different types of single and double excitations ($T^{(1)}_1$ and $T^{(1)}_2$)
\begin{align}
	\label{eq:t1_amp_tensor}
	\mathbf{t^{(1)}} = &\left\{t_{i}^{a(1)};\ t_{i}^{x(1)};\ t_{x}^{a(1)};\ t_{ij}^{ab(1)};\ t_{ij}^{ax(1)};\ t_{ix}^{ab(1)}; \right. \notag \\
	&\left. t_{ij}^{xy(1)};\ t_{xy}^{ab(1)};\ t_{ix}^{ay(1)};\ t_{ix}^{yz(1)};\ t_{xy}^{az(1)}\right\}
\end{align}
which are used to parameterize the first-order wavefunction $\ket{\Psi^{(1)}_0} = T^{(1)} \ket{\Psi_0}$ for the MR-ADC reference state. As we discussed in Ref.\@ \citenum{Chatterjee:2019p5908}, the wavefunction $\ket{\Psi^{(1)}_0}$ is equivalent to the first-order wavefunction in internally-contracted second-order $N$-electron valence perturbation theory (NEVPT2).\cite{Angeli:2001p10252,Angeli:2001p297,Angeli:2004p4043} Evaluating matrix elements of operators in \cref{eq:H_bch_0,eq:H_bch_1,eq:H_bch_2,eq:q_bch_0,eq:q_bch_1,eq:q_bch_2} also requires single- and semi-internal double-excitation amplitudes of the second-order excitation operators ($T^{(2)}_1$ and $T^{(2)}_2$):
\begin{align}
	\label{eq:t2_amp_tensor}
	\mathbf{t^{(2)}} =  \left\{t_{i}^{a(2)};\ t_{i}^{x(2)};\ t_{x}^{a(2)};\ t_{ix}^{ay(2)};\ t_{ix}^{yz(2)};\ t_{xy}^{az(2)}\right\}
\end{align}
The amplitudes $\mathbf{t^{(1)}}$ and $\mathbf{t^{(2)}}$ are determined from solving a system of projected linear equations\cite{Sokolov:2018p204113}
\begin{align}
	\label{eq:proj_amplitude_equations}
	\braket{\Psi_0|\tau_\mu\tilde{H}^{(k)}|\Psi_0} = 0 \qquad (k = 1, 2)
\end{align}
by diagonalizing small blocks of the zeroth-order Hamiltonian $H^{(0)}$ matrix in the basis of internally-contracted excitations $\tau^\dag_\mu$.\cite{Angeli:2001p10252,Angeli:2001p297,Angeli:2004p4043,Chatterjee:2019p5908} 

\begin{figure*}[t!]
	\subfloat[$\mathbf{M}$ matrix]{\label{fig:adc2_M}\includegraphics[width=0.30\textwidth]{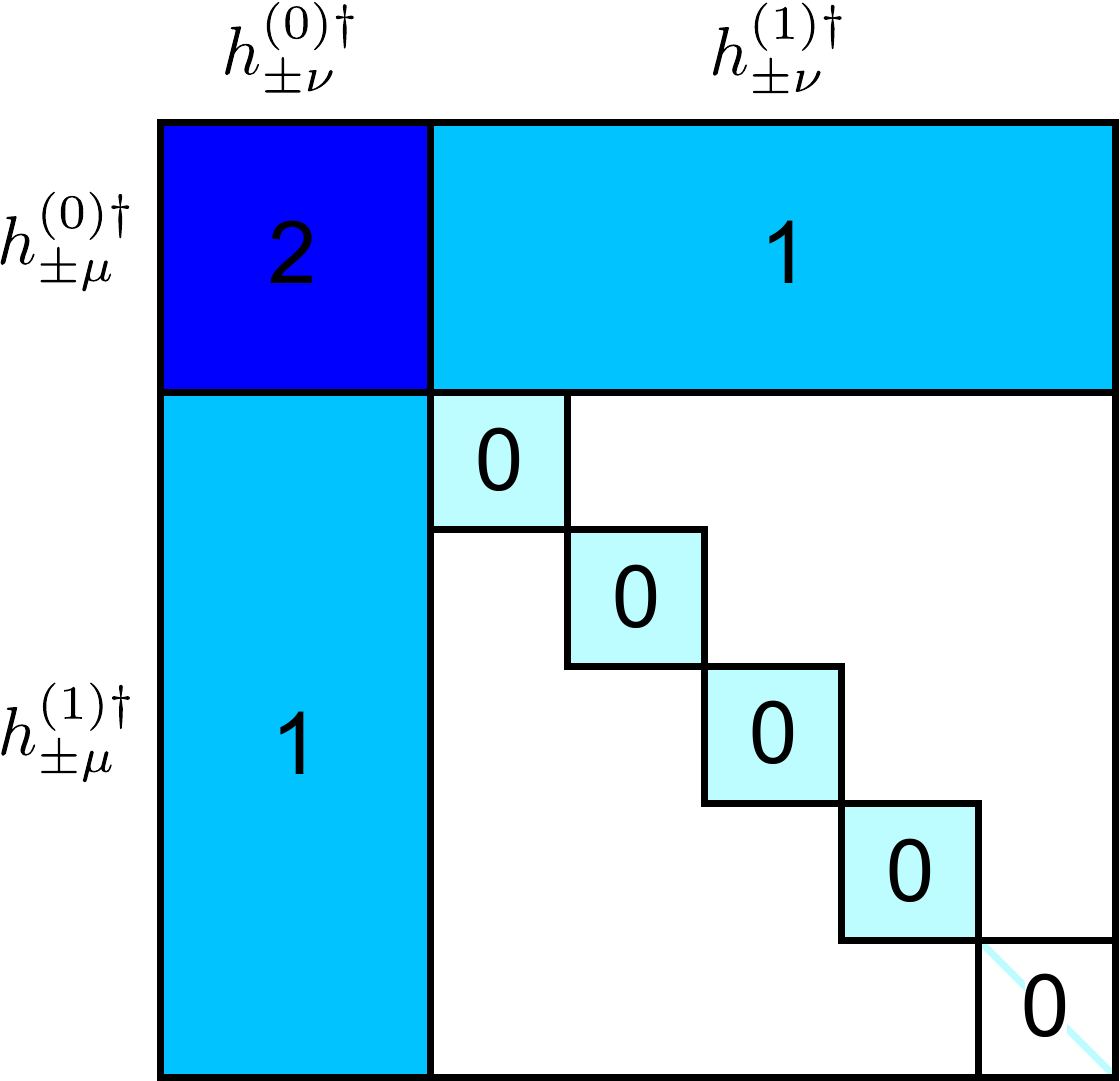}} \qquad \qquad
	\subfloat[$\mathbf{T}$ matrix]{\label{fig:adc2_T}\includegraphics[width=0.30\textwidth]{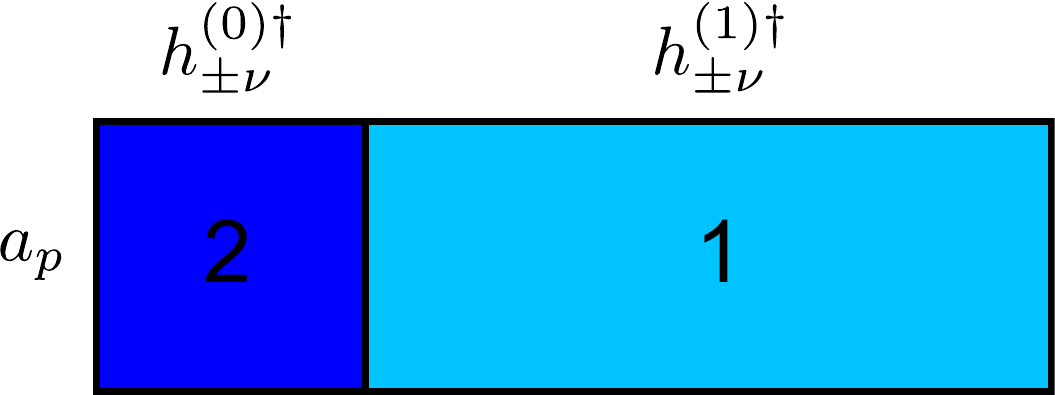}}
	\captionsetup{justification=raggedright,singlelinecheck=false}
	\caption{Perturbative structure of the effective Liouvillean ($\mathbf{M}$) and transition moments ($\mathbf{T}$) matrices of EA- and IP-MR-ADC(2). Non-zero matrix blocks are highlighted in color. A colored line represents a diagonal block. Numbers denote the perturbation order to which the effective Hamiltonian $\tilde{H}$ or observable $\tilde{q}$ operators are approximated for each block.}
	\label{fig:adc2_matrices}
\end{figure*}

\cref{fig:adc2_matrices} shows the perturbation order to which the effective Hamiltonian $\tilde{H}$ and observable $\ta{p}$ operators are expanded for each block of the EA/IP-MR-ADC(2) matrices $\mathbf{M}_\pm$ and $\mathbf{T}_\pm$. To maintain the total second order, $\ta{p}$ is expanded up to $\ta{p}^{(2)}$ and $\ta{p}^{(1)}$ for the $h_{\pm\mu}^{(0)\dagger}$ and $h_{\pm\mu}^{(1)\dagger}$ sectors of the $\mathbf{T}_\pm$ matrix, respectively (\cref{eq:T+_matrix,eq:T-_matrix}). Similarly, the diagonal $h_{\pm\mu}^{(0)\dagger}$--$h_{\pm\mu}^{(0)\dagger}$ and $h_{\pm\mu}^{(1)\dagger}$--$h_{\pm\mu}^{(1)\dagger}$ blocks of the effective Liouvillean matrix $\mathbf{M}_\pm$ are expanded up to $\tilde{H}^{(2)}$ and $\tilde{H}^{(0)}$, respectively, while the $h_{\pm\mu}^{(0)\dagger}$--$h_{\pm\mu}^{(1)\dagger}$ coupling block includes terms up to $\tilde{H}^{(1)}$. Since $\tilde{H}^{(0)}$ does not contain contributions that couple excitations outside of the active space, the $h_{\pm\mu}^{(1)\dagger}$--$h_{\pm\mu}^{(1)\dagger}$ sector of $\mathbf{M}_\pm$ has a block-diagonal structure (\cref{fig:adc2_matrices}). 
Out of five blocks of the $h_{\pm\mu}^{(1)\dagger}$--$h_{\pm\mu}^{(1)\dagger}$ sector, four non-diagonal blocks correspond to non-orthogonal excitations defined by $h_{\pm\mu}^{(1)\dagger}$ with at least one active-space index and one diagonal block corresponding to $h_{\pm\mu}^{(1)\dagger}$ with all non-active indices (i.e., $h_{+\mu}^{(1)\dagger} = a_{i}^{ab}$ for EA and $h_{-\mu}^{(1)\dagger} = a_{ij}^{a}$ for IP). For a fixed active space, the computational cost of the EA/IP-MR-ADC(2) methods is dominated by the solution of the generalized eigenvalue problem \eqref{eq:adc_eig_problem} and scales as $\mathcal{O}(M^5)$ with the size of the one-electron basis set $M$, which is equivalent to computational scaling of the single-reference EA/IP-SR-ADC(2) approximations.\cite{Banerjee:2019p224112}

\subsection{Extended EA- and IP-MR-ADC(2)}
\label{sec:theory:mr_adc_2x}
\begin{figure*}[t!]
	\subfloat[$\mathbf{M}$ matrix]{\label{fig:adc2x_M}\includegraphics[width=0.30\textwidth]{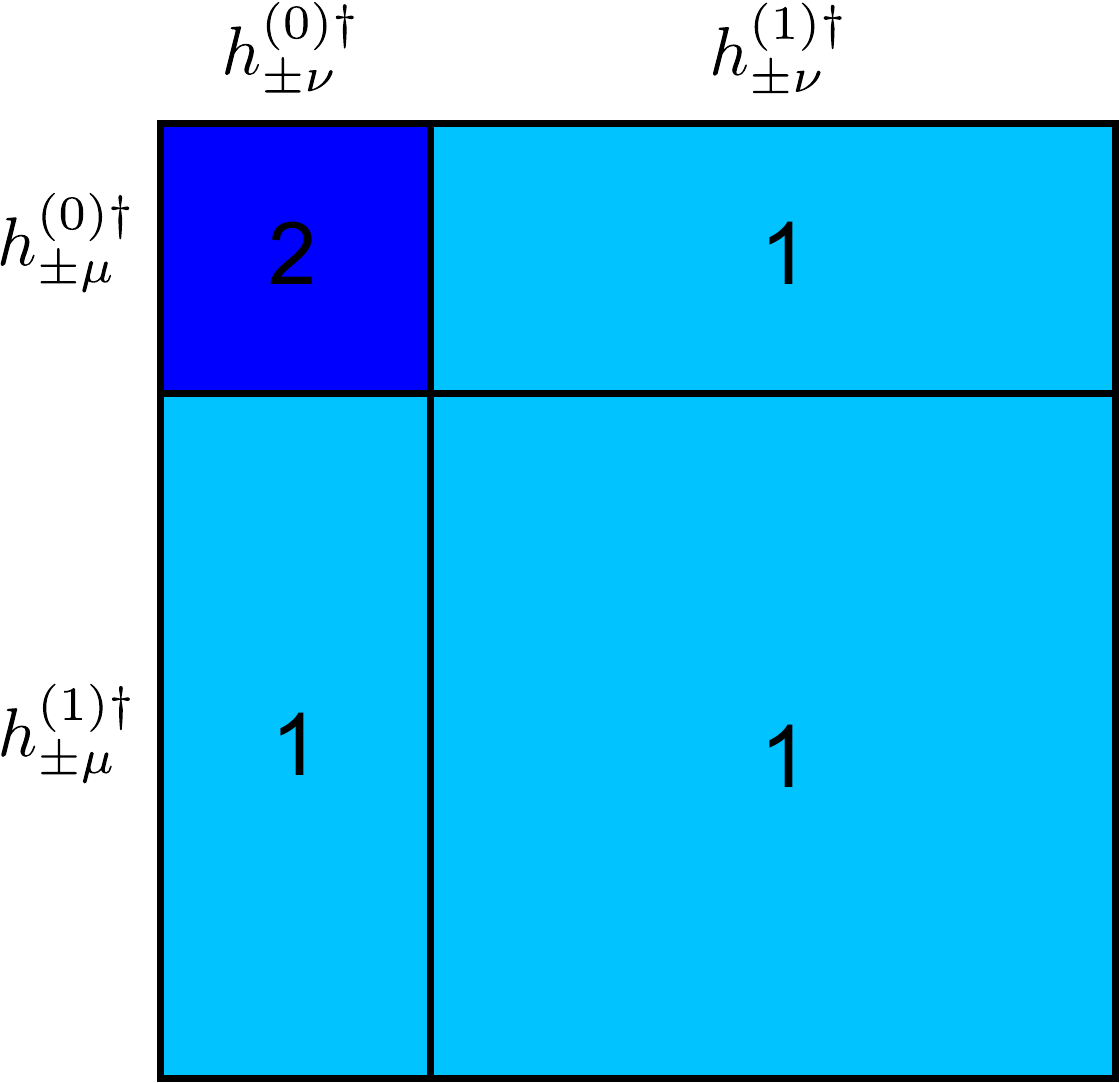}} \qquad \qquad
	\subfloat[$\mathbf{T}$ matrix]{\label{fig:adc2x_T}\includegraphics[width=0.30\textwidth]{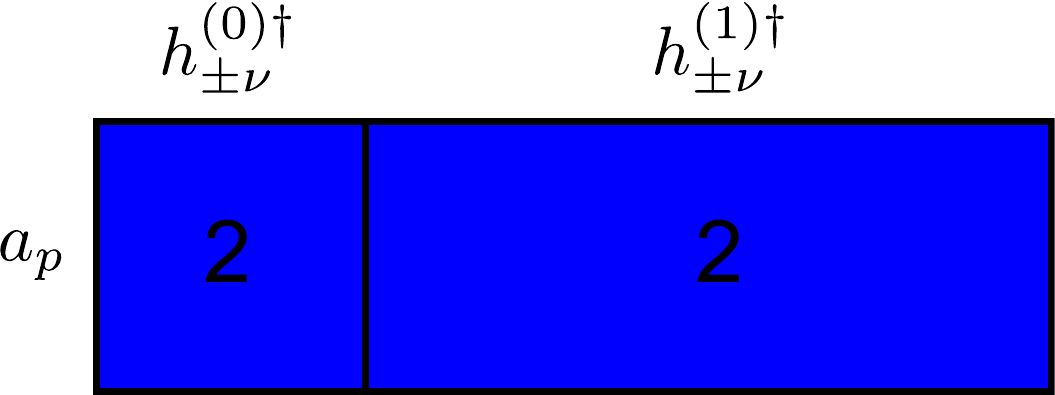}}
	\captionsetup{justification=raggedright,singlelinecheck=false}
	\caption{Perturbative structure of the effective Liouvillean ($\mathbf{M}$) and transition moments ($\mathbf{T}$) matrices of EA- and IP-MR-ADC(2)-X. Non-zero matrix blocks are highlighted in color. Numbers denote the perturbation order to which the effective Hamiltonian $\tilde{H}$ or observable $\tilde{q}$ operators are approximated for each block.}
	\label{fig:adc2x_matrices}
\end{figure*}

In addition to the EA- and IP-MR-ADC(2) approaches that include all contributions to the ADC matrices strictly to second order, in this work we also explore the extended EA- and IP-MR-ADC(2)-X approximations, which additionally incorporate contributions of $\tilde{H}^{(1)}$ in the $h_{\pm\mu}^{(1)\dagger}$--$h_{\pm\mu}^{(1)\dagger}$ block of $\mathbf{M}_\pm$ and $\ta{p}^{(2)}$ in the $h_{\pm\mu}^{(1)\dagger}$ sector of $\mathbf{T}_\pm$ (\cref{fig:adc2x_matrices}). The extended ADC approximations have been originally introduced in the single-reference ADC framework\cite{Trofimov:1995p2299,Trofimov:2005p144115,Dreuw:2014p82} as a way to partially incorporate third-order correlation effects into the description of excitation energies and transition moments. 

Increasing the order of $\tilde{H}$ in the $h_{\pm\mu}^{(1)\dagger}$--$h_{\pm\mu}^{(1)\dagger}$ block of $\mathbf{M}_\pm$ introduces couplings between excitations described by the $h_{\pm\mu}^{(1)\dagger}\ket{\Psi_0}$ basis states such that this sector of the effective Liouvillean matrix is no longer block-diagonal. These new contributions serve a two-fold purpose: (i) they significantly improve orbital relaxation effects for the primary electron-attached or ionized states described by the $h_{\pm\mu}^{(0)\dagger}$ operators, and (ii) they provide a better description of energies for the satellite transitions involving electron attachment/ionization accompanied by a single excitation outside of the active space. 

Importantly, introducing new terms in the EA- and IP-MR-ADC(2)-X equations does not increase the computational scaling of calculating transition energies with the size of the basis set or active space, relative to EA- and IP-MR-ADC(2). In particular, the $h_{\pm\mu}^{(1)\dagger}$--$h_{\pm\mu}^{(1)\dagger}$ matrix elements of the $\tilde{H}^{(1)}$ operator depend only on the first-order amplitudes $t_\mu^{(1)}$, one- and two-electron integrals, and up to three-particle reduced density matrix (3-RDM) with respect to the reference state $\ket{\Psi_0}$. Evaluating additional terms in the effective transition moments matrix $\mathbf{T}_\pm$ requires a full set of the second-order singles and doubles amplitudes
\begin{align}
	\label{eq:t2_amp_tensor_full}
	\mathbf{t^{(2)}} = &\left\{t_{i}^{a(2)};\ t_{i}^{x(2)};\ t_{x}^{a(2)};\ t_{ij}^{ab(2)};\ t_{ij}^{ax(2)};\ t_{ix}^{ab(2)}; \right. \notag \\
	&\left. t_{ij}^{xy(2)};\ t_{xy}^{ab(2)};\ t_{ix}^{ay(2)};\ t_{ix}^{yz(2)};\ t_{xy}^{az(2)}\right\}
\end{align}
which can be obtained by solving the second-order amplitude equations (\cref{eq:proj_amplitude_equations}, $k$ = 2) with the $\mathcal{O}(M^6)$ basis set scaling. 
In practice, the $\mathbf{t^{(2)}}$ amplitudes have a very small effect on the EA/IP-MR-ADC(2) and EA/IP-MR-ADC(2)-X results and their contributions can be neglected (see \cref{sec:implementation} for details), lowering computational scaling of computing transition moments to $\mathcal{O}(M^5)$.

\section{Implementation}
\label{sec:implementation}
The main objective of EA- or IP-MR-ADC is to compute charge excitation energies by numerically solving the generalized eigenvalue problem \eqref{eq:adc_eig_problem} for several (usually, low-energy) EA's or IP's and to obtain the corresponding transition probabilities by calculating the spectroscopic amplitudes in \cref{eq:spec_amplitudes}. Implementation of EA- and IP-MR-ADC(2) as well as their extended MR-ADC(2)-X variants generally follows the implementation of IP-MR-ADC(2) that we discussed in Ref. \citenum{Chatterjee:2019p5908}. Here, we briefly overview the main steps of the MR-ADC algorithm and highlight differences with our previous implementation. 

\subsection{CASSCF and CASCI Wavefunctions}
\label{sec:implementation:casscf_and_casci_wfn}
The EA- or IP-MR-ADC calculation starts by computing the $N$-electron reference CASSCF wavefunction $\ket{\Psi_0}$ for a set of core, active, and external molecular orbitals specified by the user. In addition to $\ket{\Psi_0}$, the EA/IP-MR-ADC equations require calculating the $(N\pm1)$-electron CASCI states $\ket{\Psi_{\pm I}}$ that define the active-space attachment/ionization operators $Z_{\pm I}^\dag$ in \cref{eq:ea_zeroth_order_manifold,eq:ip_zeroth_order_manifold}. While, formally, the set of the $\ket{\Psi_{\pm I}}$ states must be complete (i.e., their number scales factorially with the number of active orbitals), in practice only a small subset of low-energy $\ket{\Psi_{\pm I}}$ need to be included in the calculation. The number of these states ($N_{\mathrm{CI}}$) should be sufficiently large to include all important CASCI states in the spectral region of interest and is chosen to be a user-defined parameter in our implementation. The optimal value of $N_{\mathrm{CI}}$ can be determined by monitoring convergence of the MR-ADC results in a series of calculations with increasing number of $\ket{\Psi_{\pm I}}$.

\subsection{Reduced Density Matrices and Amplitudes of the Effective Hamiltonian}
\label{sec:implementation:rdms}
Once the CASSCF and CASCI states are computed, their wavefunctions $\ket{\Psi_0}$ and $\ket{\Psi_{\pm I}}$ are used to calculate reduced density matrices (RDMs) as expectation values of creation and annihilation operators in the active space. Three types of RDMs appear in the EA/IP-MR-ADC equations: (i) {\it reference} RDMs computed with respect to $\ket{\Psi_0}$ (e.g., $\braket{\Psi_0|\c{x}\a{y}|\Psi_0}$); (ii) {\it transition} RDMs between $\ket{\Psi_0}$ and $\ket{\Psi_{\pm I}}$ (e.g., $\braket{\Psi_{+I}|\c{x}|\Psi_0}$ or $\braket{\Psi_{0}|\c{x}|\Psi_{-I}}$); and (iii) {\it excited-state} RDMs between $\ket{\Psi_{\pm I}}$ themselves (e.g., $\braket{\Psi_{+I}|\c{x}\a{y}|\Psi_{+J}}$). Since transition RDMs are computed between states with different particle number, they always contain an odd number of creation and annihilation operators. We will refer to these RDMs as $n$.5-RDMs, where $n$.5 is obtained by dividing the total number of creation and annihilation operators by two.

The computed RDMs are used to solve equations for the amplitudes of the effective Hamiltonian $t_\mu^{(k)}$ ($k=1,2$). As we discussed in \cref{sec:theory:mr_adc_2,sec:theory:mr_adc_2x}, equations for the EA/IP-MR-ADC(2)-X transition energies depend on the single- and double-excitation $t_\mu^{(1)}$ (\cref{eq:t1_amp_tensor}) as well as the single- and semi-internal double-excitation $t_\mu^{(2)}$ (\cref{eq:t2_amp_tensor}), while a full set of $t_\mu^{(2)}$ (singles and doubles, \cref{eq:t2_amp_tensor_full}) is necessary for calculating transition moments. In practice, however, the number of terms that depend on $t_\mu^{(2)}$ in the EA/IP-MR-ADC(2)-X equations is much smaller compared to that of $t_\mu^{(1)}$ and their contributions have a minor effect on the EA/IP-MR-ADC(2)-X results. For this reason, in our implementation we approximate
\begin{align}
	\label{eq:t2_amp_tensor_approx}
	\mathbf{t^{(2)}} \approx &\left \{t_{i}^{a(2)} \right\}
\end{align}
where we neglect all $t_\mu^{(2)}$ amplitudes with active-space indices. 
As we demonstrate in the Supporting Information (Tables S1 and S2), the approximation in \cref{eq:t2_amp_tensor_approx} has a small effect on the computed transition energies ($\Omega_\mu$) and spectroscopic factors ($P$, \cref{eq:spec_factors}), with mean absolute errors of $\sim$ 0.04 and 0.01 eV for $\Omega_\mu$ of EA-MR-ADC and IP-MR-ADC, respectively, and the corresponding errors in $P$ of $\sim$ 0.0008 and 0.0003. The larger errors of this approximation for EA-MR-ADC reflect a somewhat larger sensitivity of this approximation to relaxation effects in orbitals and differential electron correlation.

Solving the EA/IP-MR-ADC(2)-X amplitude equations for $t_\mu^{(k)}$ ($k=1,2$) requires diagonalizing overlap matrices computed in the basis of internally-contracted excitations $\tau_\mu^\dag\ket{\Psi_0}$ (\cref{eq:proj_amplitude_equations}) and removing amplitudes corresponding to small overlap eigenvalues. While diagonalizing the overlap matrix can be performed very efficiently due to its block-diagonal structure, eliminating redundant semi-internal amplitudes (e.g., $t_{ix}^{ay(1)}$, $t_{ix}^{yz(1)}$, or $t_{xy}^{az(1)}$) introduces small contributions in the amplitude equations that violate size-consistency of the EA/IP-MR-ADC(2)-X energies and transition moments.\cite{Chatterjee:2019p5908} To make sure that no size-consistency-violating terms appear in the equations, we eliminate the redundant amplitudes using the approach developed by Hanauer and K\"ohn\cite{Hanauer:2012p131103} where the semi-internal excitation operators $\tau_\mu^\dag$ are transformed to a generalized normal-ordered form prior to diagonalization of the overlap matrix, which fully restores size-consistency of the MR-ADC results.

Assuming approximation for the second-order amplitudes in \cref{eq:t2_amp_tensor_approx}, the EA/IP-MR-ADC(2)-X equations formally depend on up to reference 4-RDM, transition 3.5-RDM, and excited-state 4-RDM. The only two contributions of reference 4-RDM are found in (i) the first-order equations for the semi-internal amplitudes $t_{ix}^{yz(1)}$ and $t_{xy}^{az(1)}$ and (ii) matrix elements of $\tilde{H}^{(2)}$ in the effective Liouvillean matrix $\mathbf{M}_{\pm}$. While the $t_{ix}^{yz(1)}$ and $t_{xy}^{az(1)}$ amplitudes can be computed without 4-RDM using the imaginary-time algorithm developed in our previous work,\cite{Sokolov:2018p204113,Chatterjee:2019p5908} in our present implementation we bypass 4-RDM without any approximations by forming efficient intermediates in the first-order amplitude equations, as demonstrated in the Appendix. (We note that the same technique can be used to bypass reference 4-RDM in implementation of partially-contracted state-specific and quasi-degenerate NEVPT2).\cite{Angeli:2001p10252,Angeli:2001p297,Angeli:2004p4043} We employ a similar approach to avoid calculating the remaining reference 4-RDM, transition 3.5-, and excited-state 4-RDM contributions in the equations for the $\mathbf{M}_{\pm}$ matrix elements, as outlined in Ref.\@ \citenum{Chatterjee:2019p5908}. The resulting EA/IP-MR-ADC(2)-X implementation depends on up to reference 3-RDM, transition 2.5-RDM, and excited-state 3-RDM, leading to an overall $\mathcal{O}(N_{\mathrm{det}} N^6_{\mathrm{act}})$ computational scaling of the algorithm with the number of active-space orbitals $N_{\mathrm{act}}$ and dimension of CAS Hilbert space $N_{\mathrm{det}}$.

\subsection{MR-ADC Eigenvalue Problem and Spectroscopic Factors}
\label{sec:implementation:eigen_problem}

In the final step of the EA/IP-MR-ADC(2)-X algorithm, electron attachment or ionization energies are computed by solving the generalized eigenvalue problem \eqref{eq:adc_eig_problem} transformed to the symmetrically-orthogonalized form 
\begin{align}
	\label{eq:adc_eig_problem_orthogonal}
	\mathbf{\tilde{M}}_{\pm} \mathbf{\tilde{Y}}_{\pm}  = \mathbf{\tilde{Y}}_{\pm} \boldsymbol{\Omega}_{\pm}
\end{align}
where $\mathbf{\tilde{M}}_{\pm} = \mathbf{S}^{-1/2}_{\pm} \mathbf{M}_{\pm} \mathbf{S}^{-1/2}_{\pm}$ and $\mathbf{\tilde{Y}}_{\pm} = \mathbf{S}^{1/2}_{\pm} \mathbf{Y}_{\pm}$. The eigenvectors $\mathbf{Y}_{\pm}$ are used to calculate the spectroscopic factors that provide information about the intensity of a photoelectron transition $\alpha$ with energy $\Omega_{\pm \alpha}$
\begin{align}
	\label{eq:spec_factors}
	P_{\pm \alpha} = \sum_{p} |X_{\pm p\alpha}|^2
\end{align}
where the spectroscopic amplitudes $X_{\pm p\mu}$ are defined in \cref{eq:spec_amplitudes}. Working equations for all matrix elements of $\mathbf{M}_{\pm}$, $\mathbf{T}_{\pm}$, and $\mathbf{S}_{\pm}$ in \cref{eq:adc_eig_problem_orthogonal,eq:spec_amplitudes} were automatically generated using a modified version of the \textsc{SecondQuantizationAlgebra} program\cite{Neuscamman:2009p124102} and are provided in the Supporting Information. 

Solving the eigenvalue problem \eqref{eq:adc_eig_problem_orthogonal} requires diagonalizing the overlap matrix $\mathbf{S}_{\pm}$ and removing redundant eigenvectors with small eigenvalues. Conveniently, the non-diagonal blocks of $\mathbf{S}_{\pm}$ have the same form as some of the blocks of the overlap matrix encountered in the solution of the amplitude equations (\cref{sec:implementation:rdms}), which allows to perform overlap diagonalization and truncation of the redundant eigenvectors for these two steps of the MR-ADC algorithm simultaneously.\cite{Chatterjee:2019p5908} Our implementation employs two user-defined parameters ($\eta_s$ and $\eta_d$) for discarding linearly-dependent eigenvectors of the overlap matrices. 
We use the $\eta_s$ parameter ($\sim 10^{-6}$) for eliminating redundancies in the solution of the single- and semi-internal double-excitation amplitude equations (e.g., $t_{i}^{a(1)}$, $t_{ix}^{ay(1)}$, $t_{ix}^{yz(1)}$, etc), as well as for the $\mathbf{S}_{+}$ and $\mathbf{S}_{-}$ matrix blocks with  $h_{+\mu}^{(k)\dagger} = \{ \c{a}; a_{y}^{ax} \}$ and $h_{-\mu}^{(k)\dagger} = \{ \a{i}; a_{ix}^{y} \}$, respectively. For the remaining (double-excitation) amplitudes and blocks of $\mathbf{S}_{\pm}$ that typically exhibit less severe linear dependencies, we employ a smaller truncation parameter $\eta_d$ ($\sim 10^{-10}$) to retain more eigenvectors in the overlap matrix.

The symmetric EA/IP-MR-ADC(2)-X eigenvalue problem can be solved using any available iterative eigensolver to obtain several lowest photoelectron transition energies and spectroscopic factors. In our MR-ADC program, we employ a multiroot implementation of the Davidson algorithm\cite{Davidson:1975p87,Liu:1978p49} that computes eigenvalues by starting with a set of guess (trial) eigenvectors $\mathbf{\tilde{Y}}_{\pm}$ and optimizing these vectors until convergence by forming the matrix-vector products $\boldsymbol{\sigma}_{\pm}  = \mathbf{\tilde{M}}_{\pm} \mathbf{\tilde{Y}}_{\pm}$ with the $\mathcal{O}(M^5)$ basis set scaling. To reduce the cost of forming the $\boldsymbol{\sigma}_{\pm}$ vectors, the computationally expensive but small $h_{\pm\mu}^{(0)\dagger}$--$h_{\pm\mu}^{(0)\dagger}$ block of the $\mathbf{M}_{\pm}$ matrix is precomputed at the first iteration, stored in memory, and reused in subsequent iterations of the Davidson algorithm. 

\section{Computational Details}
\label{sec:computational_details}

The EA- and IP-MR-ADC(2) methods, as well as their extended EA/IP-MR-ADC(2)-X variants, were implemented in \textsc{Prism}, a standalone program that is being developed in our group. The \textsc{Prism} program was interfaced with \textsc{PySCF}\cite{Sun:2020p024109} to obtain integrals and CASSCF/CASCI reference wavefunctions. The MR-ADC results were benchmarked against EA's and IP's from semi-stochastic heat-bath configuration interaction (SHCI)\cite{Holmes:2016p3674,Sharma:2017p1595,Holmes:2017p164111} extrapolated to full configuration interaction (FCI) limit and were compared to those computed using strongly-contracted NEVPT2 (sc-NEVPT2),\cite{Angeli:2001p10252,Angeli:2001p297} single-reference EA- and IP-ADC (EA/IP-SR-ADC(n), n = 2, 3),\cite{Banerjee:2019p224112} and equation-of-motion coupled cluster theory with single and double excitations (EA/IP-EOM-CCSD).\cite{Sinha:1989p544,Mukhopadhyay:1991p441,Nooijen:1992p55} We used \textsc{PySCF} to obtain the sc-NEVPT2 and EA/IP-SR-ADC results, while EA's and IP's from EOM-CCSD were computed using \textsc{Q-Chem}.\cite{qchem:44} The SHCI electron attachment and ionization energies were obtained by computing total energies of individual eigenstates of neutral, electron-attached, and ionized systems for a range of selection parameters and extrapolating to the FCI limit using a linear fit as described in Ref.\@ \citenum{Holmes:2017p164111}. The SHCI method was implemented in the \textsc{Dice} program.\cite{Holmes:2016p3674,Sharma:2017p1595,Holmes:2017p164111} 

Performance of the EA/IP-MR-ADC(2) and EA/IP-MR-ADC(2)-X methods was tested for a benchmark set of eight molecules (\ce{HF}, \ce{F2}, \ce{CO}, \ce{N2}, \ce{H2O}, \ce{CS}, \ce{H2CO}, and \ce{C2H4}), a carbon dimer (\ce{C2}), and an ethylene molecule in a twisted geometry (\teth). Following our previous work,\cite{Chatterjee:2019p5908} for each small molecule we considered two geometries, denoted as equilibrium and stretched. The equilibrium geometries were taken from Ref.\@ \citenum{Trofimov:2005p144115}. The stretched geometries of diatomic molecules were obtained by increasing the bond length by a factor of two. For \ce{H2O}, \ce{H2CO}, and \ce{C2H4}, the stretched geometries were defined by doubling the \ce{O-H}, \ce{C-O}, and \ce{C-C} bond distances, respectively. The C$-$C bond distance in \ce{C2} was set to 1.2425 \angstrom. The \teth geometry was obtained from Ref.\@ \citenum{Mullinax:2015p2487} and is reported in the Supporting Information.

\begin{table*}[t!]
\caption{Vertical attachment energies ($\Omega$, eV) and spectroscopic factors ($P$) of molecules with equilibrium geometries. See \cref{sec:computational_details} for details of the calculations. Also shown are mean absolute errors (\mae), standard deviations (\std), and maximum absolute errors (\maxe) of the results, relative to SHCI.}
\label{tab:ea_re}
\setstretch{1}
\scriptsize
\centering
%\begin{tabular}{L{1cm}C{1cm}C{1cm}C{1cm}C{0.1cm}C{1cm}C{1cm}C{0.1cm}C{1cm}C{1cm}C{0.1cm}C{1cm}C{0.1cm}C{0.1cm}C{2cm}C{0.1cm}C{1cm}}
\begin{threeparttable}
\begin{tabular}{lcccccccccccc}
\hline
\hline
System & State & \multicolumn{2}{c}{SR-ADC(2)} & \multicolumn{2}{c}{SR-ADC(3)} & \multicolumn{2}{c}{MR-ADC(2)} & \multicolumn{2}{c}{MR-ADC(2)-X} & EOM-CCSD & sc-NEVPT2 & SHCI\tabularnewline
 &  & $\Omega$ & $P$ & $\Omega$ & $P$ & $\Omega$ & $P$ & $\Omega$ & $P$ & $\Omega$ & $\Omega$ &$\Omega$\tabularnewline
\hline
HF 			& $4\sigma$ 		& $-0.84$	& $1.00$	& $-0.82$ 	& $0.99$	& $-0.91$	& $1.00$ 	& $-0.82$ 	& $0.99$ 	& $-0.83$ 	& $-0.88$ 	& $-0.81$\tabularnewline
 			& $5\sigma$ 		& $-4.97$	& $0.99$	& $-4.88$ 	& $0.98$	& $-5.05$	& $0.99$ 	& $-4.87$ 	& $0.98$ 	& $-4.93$	& $-4.97$ 	& $-4.88$\tabularnewline
F$_2$ 		& $3\sigma_{u}$ 	& $0.12$	& $0.91$	& $0.44$ 	& $0.89$	& $-0.41$	& $0.89$ 	& $0.19$ 	& $0.86$ 	& $0.00$ 	& $0.28$ 	& $0.27$\tabularnewline
 			& $4\sigma_{u}$ 	& $-4.84$	& $0.97$	& $-4.74$ 	& $0.96$	& $-5.01$	& $0.95$ 	& $-4.87$ 	& $0.94$ 	& $-4.84$	& $-4.68$ 	& $-4.74$\tabularnewline
CO 			& $2\pi$ 			& $-1.81$	& $0.97$	& $-1.79$ 	& $0.95$	& $-2.05$	& $0.97$ 	& $-1.84$ 	& $0.94$ 	& $-1.82$	& $-2.15$ 	& $-1.79$\tabularnewline
 			& $6\sigma$		& $-2.00$	& $0.99$	& $-1.93$ 	& $0.99$	& $-2.08$	& $1.00$ 	& $-1.99$ 	& $0.99$ 	& $-1.99$	& $-2.12$ 	& $-1.95$\tabularnewline
N$_2$ 		& $1\pi_{g}$ 		& $-2.63$	& $0.94$	& $-2.55$ 	& $0.92$	& $-2.75$	& $0.93$ 	& $-2.53$ 	& $0.91$ 	& $-2.66$	& $-2.66$ 	& $-2.59$\tabularnewline
 			& $3\sigma_{u}$ 	& $-2.62$	& $0.99$	& $-2.68$ 	& $0.99$	& $-2.77$	& $1.00$ 	& $-2.71$ 	& $0.99$ 	& $-2.65$	& $-2.69$ 	& $-2.63$\tabularnewline
H$_2$O 		& $4a_{1}$ 		& $-0.78$	& $0.99$	& $-0.75$ 	& $0.99$	& $-0.89$	& $0.99$ 	& $-0.82$ 	& $0.99$ 	& $-0.76$	& $-0.95$ 	& $-0.74$\tabularnewline
 			& $2b_{2}$ 		& $-1.51$	& $1.00$	& $-1.50$ 	& $1.00$	& $-1.57$	& $1.00$ 	& $-1.54$ 	& $0.99$ 	& $-1.50$	& $-1.64$ 	& $-1.49$\tabularnewline
CS 			& $3\pi$ 			& $-0.05$	& $0.91$	& $-0.44$ 	& $0.89$	& $0.04$	& $0.85$ 	& $-0.02$ 	& $0.86$ 	& $-0.37$	& $-0.49$ 	& $-0.40$\tabularnewline
 			& $8\sigma$ 		& $-1.47$	& $0.98$	& $-1.42$ 	& $0.98$	& $-1.65$	& $0.99$ 	& $-1.59$ 	& $0.98$ 	& $-1.48$	& $-1.52$ 	& $-1.48$\tabularnewline
H$_2$CO 	& $2b_{1}$ 		& $-1.15$	& $0.95$	& $-1.16$	& $0.93$	& $-1.32$	& $0.93$ 	& $-1.02$ 	& $0.91$ 	& $-1.24$	& $-1.60$ 	& $-1.14$\tabularnewline
 			& $6a_{1}$ 		& $-1.63$	& $0.99$	& $-1.76$ 	& $0.98$	& $-1.84$	& $0.99$ 	& $-1.72$ 	& $0.98$ 	& $-1.68$	& $-1.84$ 	& $-1.72$\tabularnewline
C$_2$H$_4$ 	& $3b_{3u}$ 		& $-1.65$	& $0.99$	& $-1.65$ 	& $0.99$	& $-1.74$	& $0.99$ 	& $-1.67$ 	& $0.99$ 	& $-1.66$	& $-1.76$ 	& $-1.66$\tabularnewline
 			& $4a_{g}$ 		& $-2.04$	& $0.95$	& $-2.10$ 	& $0.98$	& $-2.15$	& $0.93$ 	& $-1.96$ 	& $0.91$ 	& $-2.10$	& $-2.25$ 	& $-2.10$\tabularnewline
\mae 		&  				& 0.07	&  	& 0.03 	&  	& 0.20	& 	& 0.09	& 	& 0.05	& 0.14  	& \tabularnewline
\std 			&  				& 0.11	&  	& 0.05 	&  	& 0.21	& 	& 0.13	& 	& 0.07	& 0.13  	& \tabularnewline
\maxe 		&  				& 0.35	&  	& 0.18 	&  	& 0.68	& 	& 0.38	& 	& 0.27	& 0.46  	& \tabularnewline
\hline
\hline
\end{tabular}
\end{threeparttable}
\end{table*}

\begin{table*}[t!]
\caption{Vertical attachment energies ($\Omega$, eV) and spectroscopic factors ($P$) of molecules with stretched geometries. See \cref{sec:computational_details} for details of the calculations. Also shown are mean absolute errors (\mae), standard deviations (\std), and maximum absolute errors (\maxe) of the results, relative to SHCI.}
\label{tab:ea_2re}
\setstretch{1}
\scriptsize
\centering
%\begin{tabular}{L{1cm}C{1cm}C{1cm}C{1cm}C{0.1cm}C{1cm}C{1cm}C{0.1cm}C{1cm}C{1cm}C{0.1cm}C{1cm}C{0.1cm}C{0.1cm}C{2cm}C{0.1cm}C{1cm}}
\begin{threeparttable}
\begin{tabular}{lcccccccccccc}
\hline
\hline
System & State & \multicolumn{2}{c}{SR-ADC(2)} & \multicolumn{2}{c}{SR-ADC(3)} & \multicolumn{2}{c}{MR-ADC(2)} & \multicolumn{2}{c}{MR-ADC(2)-X} & EOM-CCSD & sc-NEVPT2 & SHCI\tabularnewline
 &  & $\Omega$ & $P$ & $\Omega$ & $P$ & $\Omega$ & $P$ & $\Omega$ & $P$ & $\Omega$ & $\Omega$ &$\Omega$\tabularnewline
\hline
HF 			& $4\sigma$ 				& $2.41$ 	& $0.93$	& $3.17$ 	& $0.79$	& $1.58$	& $0.79$ 	& $2.06$ 	& $0.75$ 	& $2.09$ 	& $1.80$ 	& $2.33$\tabularnewline
 			& $3\sigma$ $+$ $5\sigma$ 	& 		& 		& 	 	& 		& $-1.73$	& $0.21$ 	& $-1.75$ 	& $0.40$ 	& 		& $-1.60$ 	& $-1.49$\tabularnewline
 			& $3\sigma$ $-$ $5\sigma$	& $-1.76$	& $0.98$	& $-1.73$ 	& $0.86$& $-2.36$	& $0.82$ 	& $-2.49$ 	& $0.62$ 	& $-1.81$	& $-1.78$ 	& $-2.22$\tabularnewline
F$_2$ 		& $3\sigma_{u}$ 			& $4.06$	& $0.52$	& $8.64$	& $0.98$	& $2.87$	& $0.49$ 	& $3.58$	& $0.47$ 	& $3.92$	& $3.57$ 	& $3.80$\tabularnewline
 			& $3\sigma_{g}$ 			& 		& 		& 	 	& 		& $1.90$	& $0.43$ 	& $2.66$ 	& $0.41$ 	& 		& $3.27$ 	& $2.90$\tabularnewline
CO 			& $2\pi$ 					& $2.35$	& $0.91$	& $3.36$ 	& $0.77$	& $1.94$	& $0.56$ 	& $2.02$ 	& $0.56$ 	& $2.96$	& $1.90$ 	& $1.93$\tabularnewline
N$_2$ 		& $1\pi_{g}$ 				& $-0.65$	& $0.84$	& $13.00$ & $1.88$	& $0.33$	& $0.46$ 	& $0.41$ 	& $0.44$ 	& $0.95$	& $0.03$ 	& $0.45$\tabularnewline
H$_2$O 		& $4a_{1}$ 				& $2.35$	& $0.82$	& $2.67$ 	& $0.71$	& $1.42$	& $0.69$ 	& $1.63$ 	& $0.68$ 	& $1.33$	& $1.43$ 	& $1.59$\tabularnewline
 			& $2b_{2}$ 				& $1.58$	& $0.82$	& $2.15$ 	& $0.55$	& $0.80$	& $0.72$ 	& $1.04$ 	& $0.69$ 	& $0.68$	& $0.79$ 	& $1.03$\tabularnewline
 			& $3a_{1}$ 				& 		& 		& 	 	& 		& $0.36$	& $0.12$ 	& $0.32$ 	& $0.11$ 	& 		& $0.56$ 	& $-0.01$\tabularnewline
CS 			& $3\pi$ 					& $2.24$	& $0.90$	& $3.96$ 	& $0.41$	& $2.61$	& $0.41$ 	& $2.60$ 	& $0.41$ 	& $3.00$	& $2.44$ 	& $2.28$\tabularnewline
 			& $8\sigma$ 				& $2.66$	& $0.91$	& $2.70$ 	& $0.78$	& $1.69$	& $0.28$ 	& $1.64$ 	& $0.29$ 	& $2.50$	& $1.76$ 	& $1.47$\tabularnewline
H$_2$CO 	& $2b_{1}$ 				& $1.97$	& $0.92$	& $2.35$	& $0.80$	& $1.22$	& $0.46$ 	& $1.32$ 	& $0.46$ 	& $2.29$	& $1.37$ 	& $1.40$\tabularnewline
C$_2$H$_4$ 	& $1b_{2g}$ 				& $0.79$	& $0.80$	& $1.90$ 	& $0.63$	& $0.14$	& $0.69$ 	& $0.04$ 	& $0.69$ 	& $0.26$	& $-0.23$ 	& $-0.02$\tabularnewline
 			& $1b_{3u}$ 				& $0.73$	& $0.73$	& $1.08$ 	& $0.74$	& $0.16$	& $0.52$ 	& $0.01$ 	& $0.51$ 	& $0.21$	& $-0.31$ 	& $-0.04$\tabularnewline
\mae 		&  						& 0.59	&  	& 2.44	&  	& 0.34	& 	& 0.16	& 	& 0.51	& 0.27	& \tabularnewline
\std 			&  						& 0.58 	&  	& 3.38	&  	& 0.43	& 	& 0.20	& 	& 0.49	& 0.33	& \tabularnewline
\maxe 		&  						& 1.19	&  	& 12.55 	&  	& 1.00 	& 	& 0.33	& 	& 1.04	& 0.57  	& \tabularnewline
\hline
\hline
\end{tabular}
\end{threeparttable}
\end{table*}

All computations employed the aug-cc-pVDZ basis set\cite{Kendall:1992p6796}, with the only exception of calculations for \ce{H2CO}, \ce{C2H4}, and \teth, where the cc-pVDZ basis set was used for the hydrogen atoms. For \ce{H2CO}, \ce{C2H4}, and \teth, the SHCI computations employed the frozen-core approximation for the $1s$ orbitals. In other calculations, all electrons were correlated. Active spaces used in MR-ADC and sc-NEVPT2 are denoted as ($n$e, $m$o), where $m$ is a number of frontier molecular orbitals included in the active space and $n$ is the number of active electrons. All multireference computations in \cref{sec:results:small_molecules} included 10 active orbitals. The MR-ADC and sc-NEVPT2 calculations of electron affinities incorporated $n$ = 6 active electrons for all molecules but HF, where $n$ = 4 was used. For ionization potentials, we employed $n$ = 8, 14, 10, 10, 8, 10, 12, and 10 for \ce{HF}, \ce{F2}, \ce{CO}, \ce{N2}, \ce{H2O}, \ce{CS}, \ce{H2CO}, and \ce{C2H4}, as described in our previous work.\cite{Chatterjee:2019p5908} For \ce{C2} and \teth, the (6e, 10o) and (8e, 12o) active spaces were used, respectively. All MR-ADC calculations were performed including 10 ionized or electron-attached CASCI states in the model space. The $\eta_{s}$ = $10^{-6}$ and $\eta_{d}$ = $10^{-10}$ truncation parameters were used to eliminate redundant excitations in the solution of the MR-ADC equations (see \cref{sec:implementation:eigen_problem} for details). Throughout the manuscript, positive electron affinity implies exothermic electron attachment (i.e., EA = $E_N - E_{N+1}$), while a positive ionization energy denotes an endothermic process (IP = $E_{N-1} - E_N$).

\section{Results}
\label{sec:results}

\subsection{Benchmark: Small Molecules}
\label{sec:results:small_molecules}

\begin{table*}[t!]
\caption{Vertical ionization energies ($\Omega$, eV) and spectroscopic factors ($P$) of molecules with equilibrium geometries. See \cref{sec:computational_details} for details of the calculations. Also shown are mean absolute errors (\mae), standard deviations (\std), and maximum absolute errors (\maxe) of the results, relative to SHCI.}
\label{tab:ip_re}
\setstretch{1}
\scriptsize
\centering
%\begin{tabular}{L{1cm}C{1cm}C{1cm}C{1cm}C{0.1cm}C{1cm}C{1cm}C{0.1cm}C{1cm}C{1cm}C{0.1cm}C{1cm}C{0.1cm}C{0.1cm}C{2cm}C{0.1cm}C{1cm}}
\begin{threeparttable}
\begin{tabular}{lcccccccccccc}
\hline
\hline
System & State & \multicolumn{2}{c}{SR-ADC(2)} & \multicolumn{2}{c}{SR-ADC(3)} & \multicolumn{2}{c}{MR-ADC(2)} & \multicolumn{2}{c}{MR-ADC(2)-X} & EOM-CCSD & sc-NEVPT2 & SHCI\tabularnewline
 &  & $\Omega$ & $P$ & $\Omega$ & $P$ & $\Omega$ & $P$ & $\Omega$ & $P$ & $\Omega$ & $\Omega$ &$\Omega$\tabularnewline
\hline
HF 			& $1\pi$ 			& 14.41 	& 0.89 	& 16.79 	& 0.93 	& 16.35	& 0.93 	& 16.27 	& 0.93 	& 15.85 	& 16.41 	& 16.07\tabularnewline
 			& $3\sigma$ 		& 18.69 	& 0.90 	& 20.65 	& 0.94 	& 20.38	& 0.94 	& 20.30 	& 0.93 	& 19.88	& 20.43 	& 20.06\tabularnewline
F$_2$ 		& $1\pi_{g}$ 		& 13.90 	& 0.87 	& 16.03 	& 0.89 	& 16.55	& 0.88 	& 16.01 	& 0.88 	& 15.40 	& 15.47 	& 15.64\tabularnewline
 			& $1\pi_{u}$ 		& 17.06 	& 0.84 	& 19.25 	& 0.75 	& 19.86	& 0.80 	& 18.39 	& 0.77 	& 18.77	& 18.66 	& 18.83\tabularnewline
 			& $3\sigma_{g}$ 	& 20.25 	& 0.89 	& 21.26 	& 0.89 	& 22.08	& 0.87 	& 21.95 	& 0.86 	& 21.16	& 20.91 	& 21.15\tabularnewline
CO 			& $5\sigma$ 		& 13.78 	& 0.91 	& 13.57 	& 0.90 	& 14.07	& 0.92 	& 13.84 	& 0.91 	& 13.99	& 13.46 	& 13.74\tabularnewline
 			& $1\pi$ 			& 16.24 	& 0.89 	& 17.16 	& 0.90 	& 17.38	& 0.90 	& 17.22 	& 0.90 	& 16.93	& 16.71 	& 16.90\tabularnewline
 			& $4\sigma$ 		& 18.28 	& 0.85 	& 20.46 	& 0.76 	& 20.15	& 0.85 	& 19.97 	& 0.84 	& 19.67	& 19.43 	& 19.56\tabularnewline
N$_2$ 		& $3\sigma_{g}$ 	& 14.79 	& 0.88 	& 15.42 	& 0.91 	& 15.76	& 0.91 	& 15.54 	& 0.90 	& 15.43	& 15.24 	& 15.30\tabularnewline
 			& $1\pi_{u}$ 		& 16.98 	& 0.91 	& 16.60 	& 0.92 	& 17.33	& 0.92 	& 17.17 	& 0.92 	& 17.11	& 16.76 	& 16.83\tabularnewline
 			& $2\sigma_{u}$ 	& 17.96 	& 0.85 	& 18.79 	& 0.82 	& 19.00	& 0.83 	& 18.81 	& 0.82 	& 18.71	& 18.43 	& 18.50\tabularnewline
H$_2$O 		& $1b_{1}$ 		& 11.23 	& 0.89 	& 12.99 	& 0.92 	& 12.74	& 0.93 	& 12.64 	& 0.92 	& 12.38	& 12.49 	& 12.53\tabularnewline
 			& $3a_{1}$ 		& 13.53 	& 0.89 	& 15.28 	& 0.92 	& 15.07	& 0.93 	& 14.99 	& 0.92 	& 14.66	& 14.81 	& 14.81\tabularnewline
 			& $1b_{2}$ 		& 17.95 	& 0.90 	& 19.34 	& 0.93 	& 19.28	& 0.94 	& 19.18 	& 0.93 	& 18.89	& 19.01 	& 18.98\tabularnewline
CS 			& $7\sigma$ 		& 10.99 	& 0.86 	& 10.99 	& 0.85 	& 11.59	& 0.85 	& 11.30 	& 0.84 	& 11.36	& 10.94 	& 11.13\tabularnewline
 			& $2\pi$ 			& 12.84 	& 0.91 	& 12.67 	& 0.90 	& 13.43	& 0.91 	& 13.20 	& 0.90 	& 12.94	& 12.77 	& 12.83\tabularnewline
 			& $6\sigma$ 		& 16.88 	& 0.85 	& 15.53 	& 0.18 	& 16.83	& 0.40 	& 16.55 	& 0.36 	& 17.02	& 15.79 	& 15.88\tabularnewline
H$_2$CO 	& $2b_{2}$ 		& 9.46 	& 0.87 	& 11.11 	& 0.91 	& 11.23	& 0.92 	& 10.93 	& 0.90 	& 10.62	& 10.29 	& 10.72\tabularnewline
 			& $1b_{1}$ 		& 13.73 	& 0.88 	& 14.54 	& 0.88 	& 15.14	& 0.90 	& 14.86 	& 0.89 	& 14.47	& 14.09 	& 14.48\tabularnewline
 			& $5a_{1}$ 		& 14.62 	& 0.86 	& 16.61 	& 0.90 	& 16.70	& 0.90 	& 16.39 	& 0.89 	& 15.95	& 15.68 	& 16.01\tabularnewline
 			& $1b_{2}$ 		& 16.67 	& 0.88 	& 17.04 	& 0.69 	& 17.76	& 0.88 	& 17.26 	& 0.86 	& 17.21	& 16.58 	& 16.86\tabularnewline
C$_2$H$_4$ 	& $1b_{1u}$ 		& 10.14 	& 0.91 	& 10.47 	& 0.91 	& 11.01	& 0.90 	& 10.80 	& 0.89 	& 10.58	& 10.58 	& 10.58\tabularnewline
 			& $1b_{1g}$ 		& 12.79 	& 0.91 	& 13.22 	& 0.91 	& 13.75	& 0.92 	& 13.45 	& 0.90 	& 13.22	& 13.09 	& 13.21\tabularnewline
 			& $3a_{g}$ 		& 13.78 	& 0.89 	& 14.34 	& 0.91 	& 14.74	& 0.89 	& 14.37 	& 0.87 	& 14.31	& 14.24 	& 14.25\tabularnewline
 			& $1b_{2u}$ 		& 16.13 	& 0.87 	& 16.50 	& 0.74 	& 17.10	& 0.84 	& 16.81 	& 0.83 	& 16.61	& 16.51 	& 16.45\tabularnewline
\mae 		&  				& 0.83 	&  		&  0.30	&  		& 0.56 	&  		& 0.31	& 		& 0.17	& 0.17   	& \tabularnewline
\std 			&  				& 0.68 	&  		&  0.32	&  		& 0.23  	&  		& 0.22	& 		& 0.28	& 0.19   	& \tabularnewline
\maxe 		&  				& 1.78	&  		&  0.91	&  		& 1.03	& 		& 0.80	& 		& 1.14	& 0.44  	& \tabularnewline
\hline
\hline
\end{tabular}
\end{threeparttable}
\end{table*}

\begin{table*}[t!]
\caption{Vertical ionization energies ($\Omega$, eV) and spectroscopic factors ($P$) of molecules with stretched geometries. See \cref{sec:computational_details} for details of the calculations. Also shown are mean absolute errors (\mae), standard deviations (\std), and maximum absolute errors (\maxe) of the results, relative to SHCI.}
\label{tab:ip_2re}
\setstretch{1}
\scriptsize
\centering
%\begin{tabular}{L{1cm}C{1cm}C{1cm}C{1cm}C{0.1cm}C{1cm}C{1cm}C{0.1cm}C{1cm}C{0.1cm}C{0.1cm}C{2cm}C{0.1cm}C{1cm}}
\begin{tabular}{lcccccccccccc}
\hline
\hline
System & State & \multicolumn{2}{c}{SR-ADC(2)} & \multicolumn{2}{c}{SR-ADC(3)} & \multicolumn{2}{c}{MR-ADC(2)} & \multicolumn{2}{c}{MR-ADC(2)-X} & EOM-CCSD & sc-NEVPT2 & SHCI\tabularnewline
 &  & $\Omega$ & $P$ & $\Omega$ & $P$ & $\Omega$ & $P$ & $\Omega$ & $P$ & $\Omega$ & $\Omega$ & $\Omega$ \tabularnewline
\hline
HF 				& $1\pi$ 			& 9.84 	& 0.77 	& 16.15 	& 0.84 	& 13.86	& 0.60	& 13.80	& 0.60	& 13.67 	& 13.60 	& 13.65\tabularnewline
				& $3\sigma$ 		& 13.30 	& 0.84 	& 14.68 	& 0.76 	& 14.98	& 0.73	& 14.92	& 0.72	& 14.76	& 14.83 	& 14.84\tabularnewline
F$_{2}$ 			& $1\pi_{g}$ 		& 10.63 	& 0.64 	& 17.55 	& 0.88 	& 18.12	& 0.74	& 17.46	& 0.73	& 16.86	& 17.03 	& 17.13\tabularnewline
				& $1\pi_{u}$ 		& 10.66 	& 0.64 	& 17.69 	& 0.89 	& 18.16	& 0.82	& 17.52	& 0.82	& 16.95	& 17.18 	& 17.19\tabularnewline
N$_{2}$ 			& $3\sigma_{g}$ 	& 15.70 	& 0.63 	& $-$2.60 	& 1.69 	& 14.00	& 0.69	& 13.65	& 0.68	& 14.36	& 13.10 	& 13.38\tabularnewline
				& $1\pi_{u}$ 		& 17.50 	& 0.55 	& $-$5.24 	& 2.16 	& 14.17	& 0.51	& 13.88	& 0.50	& 14.77	& 13.19 	& 13.49\tabularnewline
H$_{2}$O 			& $1b_{1}$ 		& 6.53 	& 0.71 	& 12.24 	& 0.66 	& 11.31	& 0.64	& 11.22	& 0.64	& 10.65	& 10.99 	& 11.07\tabularnewline
				& $3a_{1}$ 		& 10.49 	& 0.75 	& 12.78 	& 0.67 	& 13.22	& 0.67	& 13.14	& 0.67	& 12.69	& 13.00 	& 13.02\tabularnewline
				& $1b_{2}$ 		& 11.18 	& 0.75 	& 13.01 	& 0.72 	& 13.78	& 0.71	& 13.69	& 0.71	& 13.26	& 13.55 	& 13.56\tabularnewline
H$_{2}$CO 		& $2b_{2}$ 		& 10.65 	& 0.85 	& 8.31 	& 0.21 	& 11.51	& 0.39	& 10.96	& 0.31	& 9.85	& 10.10 	& 10.37\tabularnewline
 				& $1b_{1}$		& 10.69 	& 0.86 	& 8.35 	& 0.22 	& 11.21	& 0.48	& 11.01	& 0.47	& 9.66	& 10.27 	& 10.55\tabularnewline
 				& $5a_{1}$ 		& 10.60 	& 0.91 	& 10.97 	& 0.88 	& 13.16	& 0.57	& 12.91	& 0.57	& 10.97	& 12.83 	& 13.16\tabularnewline
C$_{2}$H$_{4}$ 	& $1b_{1u}$ 		& 9.37 	& 0.76 	& 6.87 	& 0.83 	& 9.69	& 0.53	& 9.18	& 0.52	& 9.41	& 9.15	& 9.25\tabularnewline
				& $3a_{g}$ 		& 11.38 	& 0.79 	& 8.74 	& 0.91 	& 11.36	& 0.73	& 11.14	& 0.68	& 11.17	& 10.14 	& 10.93\tabularnewline
\mae				&  				& 2.70  	&  		& 3.66 	&  		& 0.50	&  		& 0.25	& 		& 0.56 	& 0.19	& \tabularnewline
\std 				&  				& 3.10  	&  		& 6.28 	&  		& 0.36	&  		& 0.22	& 		& 0.81	& 0.21	& \tabularnewline
\maxe 			&  				& 6.53	&  		& 18.73 	&  		& 1.14	& 		& 0.59	& 		& 2.18	& 0.79		& \tabularnewline
\hline
\hline
\end{tabular}
\end{table*}

We now analyze performance of EA/IP-MR-ADC(2) and EA/IP-MR-ADC(2)-X by comparing results of these methods with accurate electron affinities (EA's) and ionization potentials (IP's) computed at the full configuration interaction (FCI) limit using the semi-stochastic heat-bath CI algorithm (SHCI). We first consider a set of eight closed-shell molecules (\ce{HF}, \ce{F2}, \ce{CO}, \ce{N2}, \ce{H2O}, \ce{CS}, \ce{H2CO}, and \ce{C2H4}) and benchmark the MR-ADC methods together with the single-reference ADC approximations (EA/IP-SR-ADC(n), n = 2 and 3), equation-of-motion coupled cluster theory with single and double excitations (EA/IP-EOM-CCSD), as well as strongly-contracted second-order N-electron valence perturbation theory (sc-NEVPT2). 

\begin{figure*}[t!]
    \subfloat[]{\label{fig:ea_mae_std_re}\includegraphics[width=0.43\textwidth]{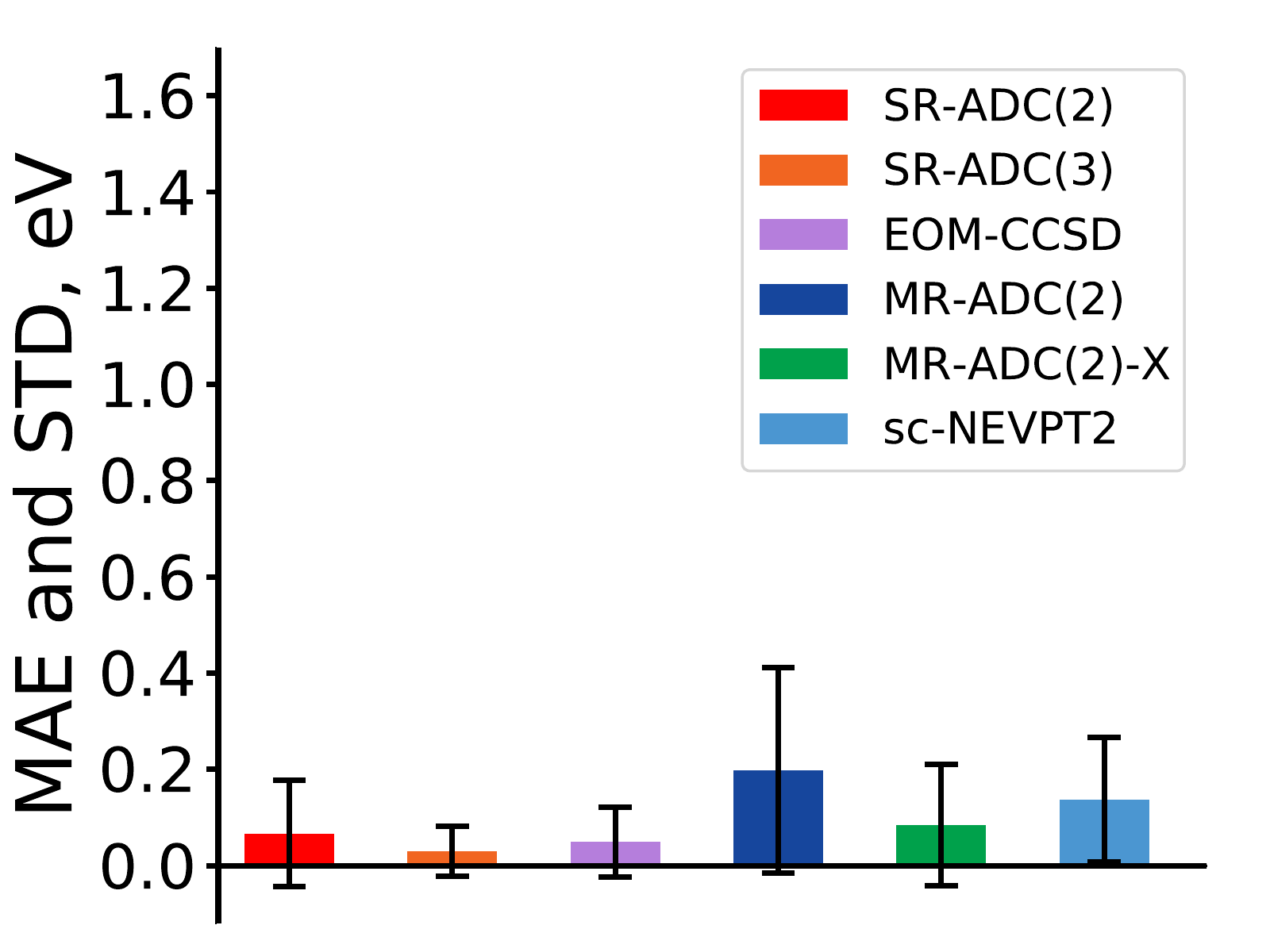}} \quad
    \subfloat[]{\label{fig:ea_mae_std_2re}\includegraphics[width=0.43\textwidth]{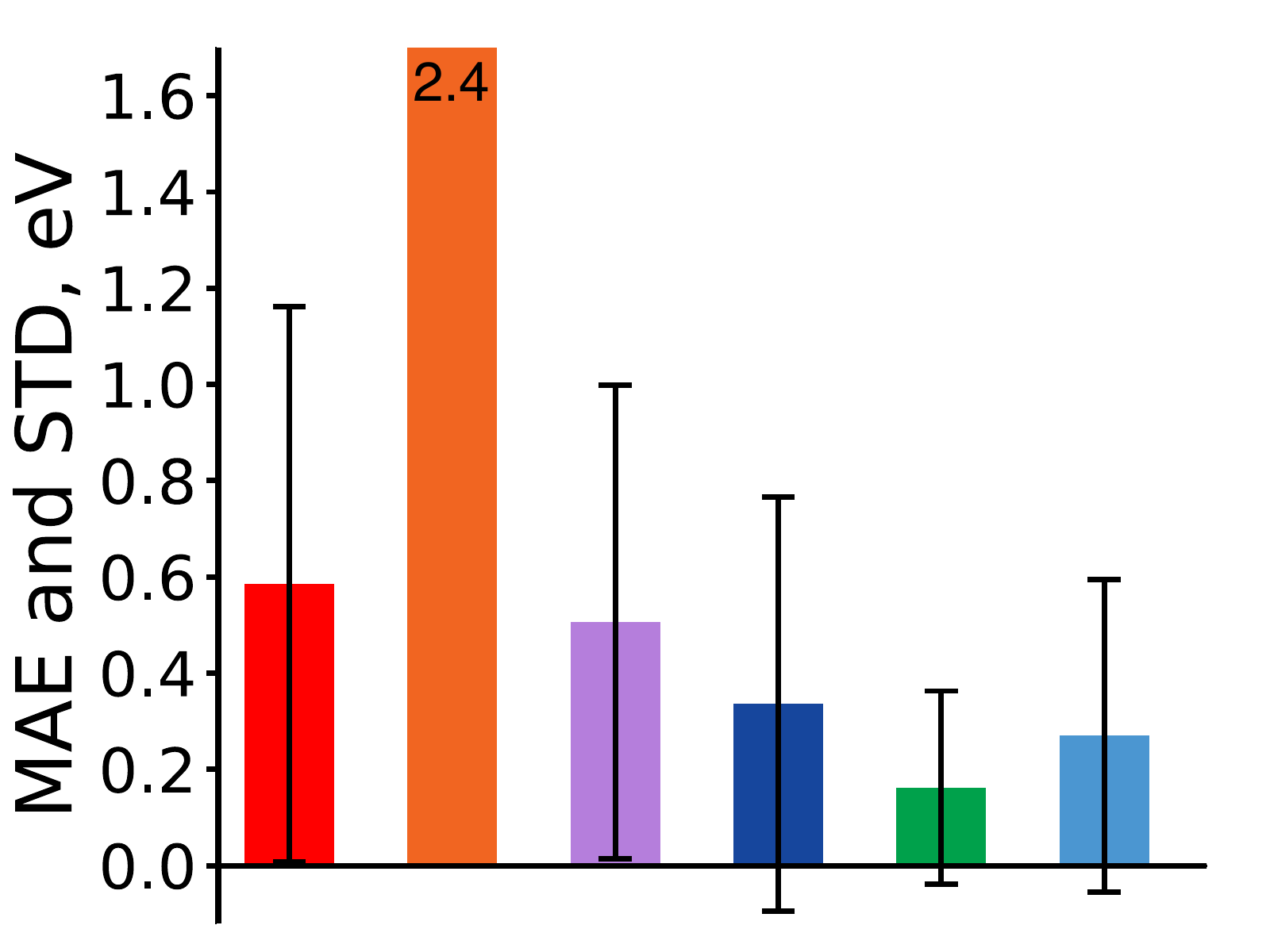}} 
    \captionsetup{justification=raggedright,singlelinecheck=false}
    \caption{Mean absolute errors (MAE, eV) and standard deviations from the mean signed error (STD, eV) for vertical electron attachment energies of molecules with (a) equilibrium and (b) stretched geometries, relative to SHCI. MAE is represented as a height of each colored bar, while STD is depicted as a radius of the black vertical line. A number on a bar indicates MAE for a bar off the chart. See \cref{tab:ea_re,tab:ea_2re} for data on individual molecules. }
    \label{fig:ea_mae_std}
\end{figure*}

\begin{figure*}[t!]
    \subfloat[]{\label{fig:ip_mae_std_re}\includegraphics[width=0.43\textwidth]{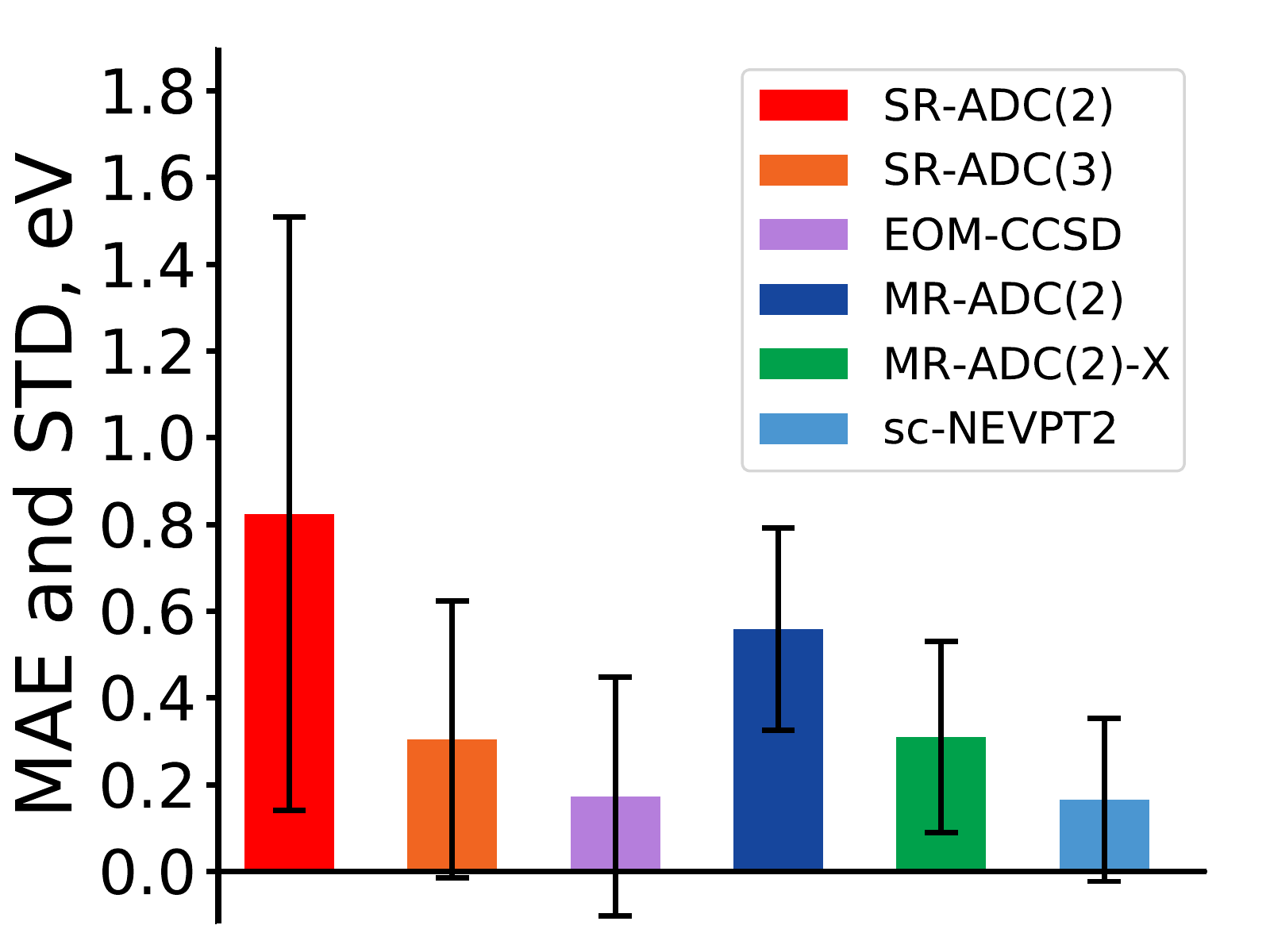}} \quad
    \subfloat[]{\label{fig:ip_mae_std_2re}\includegraphics[width=0.43\textwidth]{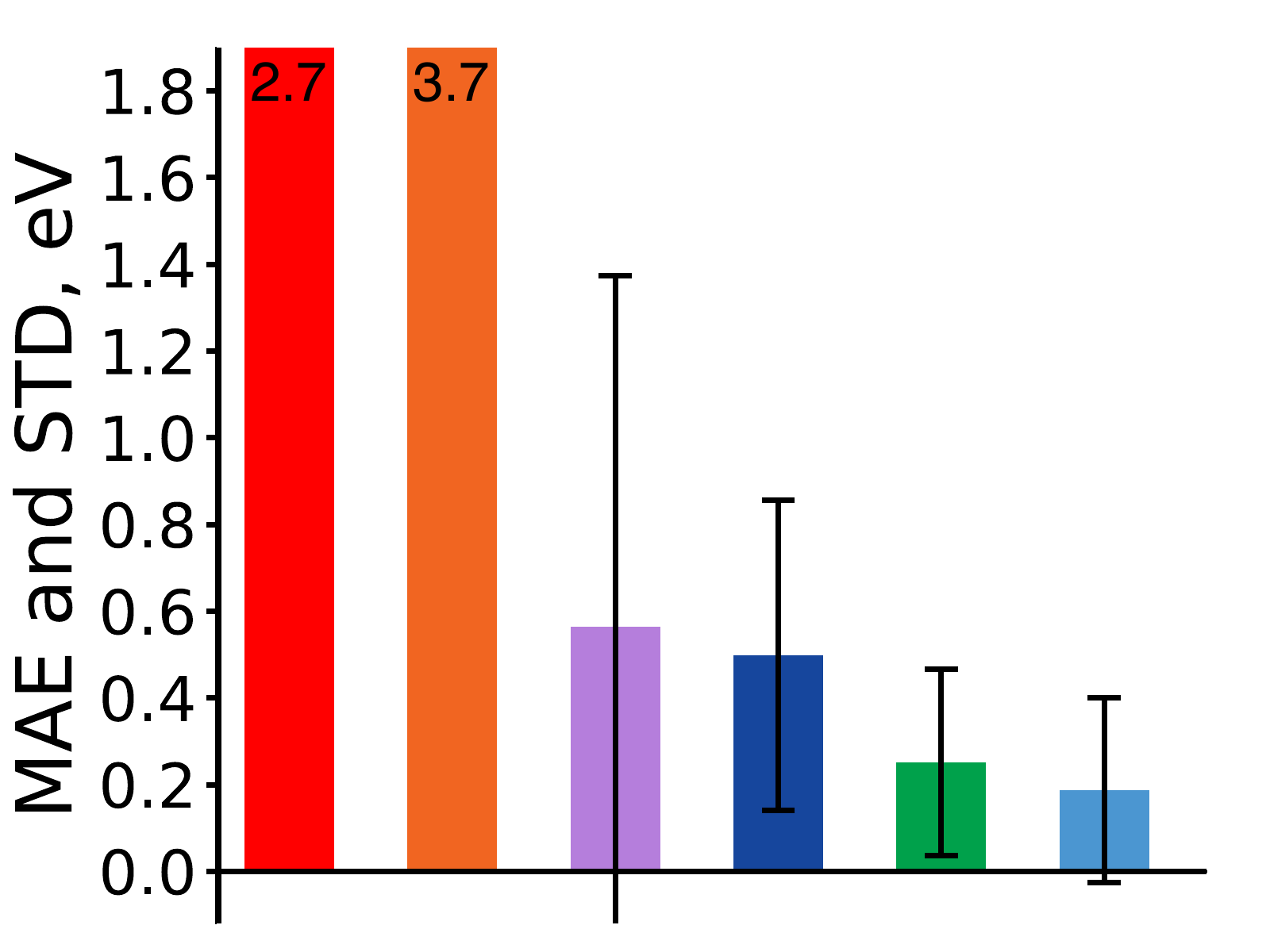}}
    \captionsetup{justification=raggedright,singlelinecheck=false}
    \caption{Mean absolute errors (MAE, eV) and standard deviations from the mean signed error (STD, eV) for vertical ionization energies of molecules with (a) equilibrium and (b) stretched geometries, relative to SHCI. MAE is represented as a height of each colored bar, while STD is depicted as a radius of the black vertical line. A number on a bar indicates MAE for a bar off the chart. See \cref{tab:ip_re,tab:ip_2re} for data on individual molecules.}
    \label{fig:ip_mae_std}
\end{figure*}

\cref{tab:ea_re} compares results of six approximate methods with SHCI for the first two EA's of small molecules computed at near-equilibrium geometries. For all molecules but \ce{F2} the computed EA's are negative, indicating that electron attachment is endothermic. All single- and multireference methods show a very good agreement with SHCI vertical electron affinities with mean absolute errors (\mae) of $\le$ 0.2 eV. Among the multireference approaches, the largest errors are produced by EA-MR-ADC(2) with \mae = 0.20 eV and a standard deviation (\std) of 0.21 eV. Including the third-order correlation effects in the EA-MR-ADC(2)-X method reduces \mae by a factor of two (\mae = 0.09 eV) and significantly lowers the standard deviation (\std = 0.13 eV). The sc-NEVPT2 method shows intermediate performance between that of EA-MR-ADC(2) and EA-MR-ADC(2)-X with \mae = 0.14 eV and \std = 0.13 eV. \cref{tab:ea_re} also shows spectroscopic factors ($P$) computed using the SR-ADC and MR-ADC approximations that provide information about the occupancy of virtual states probed by electron attachment. With the exception of just two states ($3\sigma_{u}$ of \ce{F2} and $3\pi$ of \ce{CS}), the values of spectroscopic factors are greater than 0.9, suggesting that the corresponding virtual states are largely unoccupied in the neutral molecules. For all transitions, the $P$ values computed using EA-MR-ADC(2) and EA-MR-ADC(2)-X are in closer agreement with SR-ADC(3) than SR-ADC(2).

We now turn our attention to \cref{tab:ea_2re}, which presents EA's of small molecules computed at stretched geometries where multireference effects become significant. The importance of static correlation at these geometries is demonstrated by the divergence of the SR-ADC perturbation series with more than a four-fold increase of \mae from EA-SR-ADC(2) (0.59 eV) to EA-SR-ADC(3) (2.44 eV) and a ten-fold increase in \mae for EA-EOM-CCSD (0.51 eV) relative to that for the equilibrium geometries. The best agreement with SHCI is demonstrated by EA-MR-ADC(2)-X (\mae = 0.16 eV, \std = 0.20 eV) that significantly improves on the performance of sc-NEVPT2 (\mae = 0.27 eV, \std = 0.33 eV) and EA-MR-ADC(2) (\mae = 0.34 eV, \std = 0.43 eV). Interestingly, stretching geometries of small molecules significantly depopulates their highest-occupied molecular orbitals (HOMO's), making electron attachment to these orbitals possible. While single-reference methods do not predict electron attachment to HOMO, such transitions are observed in the SHCI, EA-MR-ADC, and sc-NEVPT2 results for HF ($3\sigma$), \ce{F2} ($3\sigma_{g}$), and \ce{H2O} ($3a_{1}$). In particular, for the HF molecule electron attachment to HOMO ($3\sigma$) and the second unoccupied virtual orbital ($5\sigma$) are energetically nearly-degenerate, which results in strong mixing of the wavefunctions for these two electronic states (denoted as $3\sigma \pm 5 \sigma$ in \cref{tab:ea_2re}) as observed in the SHCI calculations. Analysis of contributions to the EA-MR-ADC(2) and EA-MR-ADC(2)-X spectroscopic factors for the \ce{HF} molecule also reveals the mixed nature of the $3\sigma \pm 5 \sigma$ transitions, in a good agreement with the SHCI results. 

To assess performance of the MR-ADC methods for ionization energies, we consider results of IP-MR-ADC(2) and IP-MR-ADC(2)-X for molecules at equilibrium and stretched geometries presented in \cref{tab:ip_re,tab:ip_2re}, respectively.  As discussed in our previous work,\cite{Chatterjee:2019p5908} IP-MR-ADC(2) shows intermediate performance between IP-SR-ADC(2) and IP-SR-ADC(3) at equilibrium (\mae = 0.56 eV and \std = 0.23 eV), but is much more reliable than SR-ADC at stretched geometries, where it maintains its accuracy (\mae = 0.50 eV and \std = 0.36 eV) while results of the single-reference ADC methods drastically deteriorate (\mae and \std $>$ 2 eV). The extended IP-MR-ADC(2)-X approximation significantly improves on IP-MR-ADC(2) lowering \mae by about a factor of two for both equilibrium and stretched geometries (\mae $\sim$ 0.3 eV and \std $\sim$ 0.2 eV). At equilibrium, the accuracy of IP-MR-ADC(2)-X is similar to IP-SR-ADC(3) (\mae and \std $\sim$ 0.3 eV) and is competitive with that of IP-EOM-CCSD (\mae $\sim$ 0.2 eV and \std $\sim$ 0.3 eV) and sc-NEVPT2 (\mae and \std $\sim$ 0.2 eV). For stretched geometries, performance of IP-MR-ADC(2)-X is quite similar to sc-NEVPT2. Both IP-MR-ADC(2) and IP-MR-ADC(2)-X yield very similar spectroscopic factors that significantly reduce their values upon stretching, indicating significant depopulation of the corresponding occupied orbitals.

Overall, our results demonstrate a consistent performance of the MR-ADC(2) and MR-ADC(2)-X approximations for electron attachment and ionization energies of small molecules across two sets of molecular geometries, as depicted in \cref{fig:ea_mae_std,fig:ip_mae_std}. Importantly, incorporating third-order effects in the MR-ADC(2)-X effective Hamiltonian matrix significantly reduces errors in computed EA's and IP's without increasing the overall computational scaling of the method. The accuracy of IP-MR-ADC(2)-X is similar to sc-NEVPT2 for transition energies, but the former method has an added advantage of providing efficient access to spectroscopic factors, which can be used to compute density of states and intensities in photoelectron spectra.

\subsection{Carbon Dimer}
\label{sec:results:carbon_dimer}

Next, we apply MR-ADC(2) and MR-ADC(2)-X to calculate low-energy EA's and IP's of the carbon dimer (\ce{C2}) at near-equilibrium geometry (1.2425 \angstrom) and compare their results with SR-ADC(3), sc-NEVPT2, and SHCI. We employ the aug-cc-pVDZ basis set in all calculations and the (6e, 10o) active space for the multireference methods. Electronic structure of the neutral \ce{C2} molecule is a well-known challenging multireference problem\cite{Roos:1987p399,Bauschlicher:1987p1919,Watts:1992p6073,Abrams:2004p9211,Wouters:2014p1501,Holmes:2017p164111} that has a closed-shell $X^1\Sigma_g^+$ ground electronic state [$(2\sigma_u)^2(1\pi_u)^4(3\sigma_g)^0$ electronic configuration with a significant contribution from $(2\sigma_u)^0(1\pi_u)^4(3\sigma_g)^2$] and several low-lying excited states with different occupations in its frontier $1\pi_u$ and $3\sigma_g$ molecular orbitals. Due to the close proximity of the $2\sigma_u$, $1\pi_u$, and $3\sigma_g$ orbital energies, the photoelectron spectrum of \ce{C2} exhibits several closely-spaced transitions with intense (primary) peaks, originating from attachment/ionization of a single electron in the neutral ground-state $(2\sigma_u)^2(1\pi_u)^4(3\sigma_g)^0$ configuration, and weak (satellite) peaks, which involve a one-electron attachment/ionization and a simultaneous single or double excitation. 

\begin{table*}[t!]
\caption{Carbon dimer vertical electron attachment (C$_2^-$) and ionization (C$_2^+$) energies ($\Omega$, eV) and spectroscopic factors ($P$) computed using the aug-cc-pVDZ basis set with $r$\ce{(C-C)} = 1.2425 \angstrom. For MR-ADC and sc-NEVPT2, the CASSCF reference wavefunction was computed using the (6e, 10o) active space. Also shown are mean absolute errors (\mae) and standard deviations (\std) for relative energies of primary (singly-ionized) and satellite states, relative to SHCI.}
\label{tab:c2}
\setstretch{1}
\scriptsize
\centering
%\begin{tabular}{L{1cm}C{1cm}C{1cm}C{1cm}C{0.1cm}C{1cm}C{1cm}C{0.1cm}C{1cm}C{0.1cm}C{0.1cm}C{2cm}C{0.1cm}C{1cm}}
\begin{threeparttable}
\begin{tabular}{lcccccccccc}
\hline
\hline
Molecule & Configuration & State & \multicolumn{2}{c}{SR-ADC(3)} & \multicolumn{2}{c}{MR-ADC(2)} & \multicolumn{2}{c}{MR-ADC(2)-X} & sc-NEVPT2 & SHCI\tabularnewline
  &  & & $\Omega$ & $P$ & $\Omega$ & $P$ & $\Omega$ & $P$ & $\Omega$ & $\Omega$\tabularnewline
\hline
C$_2^-$ 	& $(2\sigma_u)^2(1\pi_u)^4(3\sigma_g)^1$		& $1^2\Sigma_g^+$	& 3.97	& 0.8126	& 3.50	& 0.7105	& 3.29	& 0.7335	& 2.72	& 3.03 \tabularnewline
		& $(2\sigma_u)^1(1\pi_u)^4(3\sigma_g)^2$		& $1^2\Sigma_u^+$	& 2.58	& 0.0259	& 1.42	& 0.0882	& 1.09	& 0.0921	& 0.36	& 0.75 \tabularnewline
		& $(2\sigma_u)^2(1\pi_u)^4(3\sigma_u)^1$ 		& $2^2\Sigma_u^+$	& $-$1.68	& 0.9242	& $-$2.03	& 0.9882	& $-$2.01	& 0.9845	& $-$1.86	& $-$1.69 \tabularnewline
		& $(2\sigma_u)^2(1\pi_u)^4(4\sigma_g)^1$		& $2^2\Sigma_g^+$	& $-$2.78	& 0.8515	& $-$2.76	& 0.9203	& $-$2.78	& 0.9249	& $-$2.88	& $-$2.42 \tabularnewline
		& $(2\sigma_u)^2(1\pi_u)^4(1\pi_g)^1$			& $1^2\Pi_g$		& $-$2.50	& 0.9091	& $-$2.30	& 0.8601	& $-$2.36	& 0.8842	& $-$2.33	& $-$2.60 \tabularnewline
\hline
C$_2^+$ 	& $(2\sigma_u)^2(1\pi_u)^3$					& $1^2\Pi_u$		& 11.69 		& 0.9215 		& 12.33  	& 0.8907	& 12.03	& 0.8747	& 12.32 	& 12.34\tabularnewline
		& $(2\sigma_u)^2(1\pi_u)^2(3\sigma_g)^1$		& $1^2\Delta_g$	& 11.17 		& 0.0002 		& 14.51 	& 0.0003	& 14.36	& 0.0003	& 13.99 	& 13.94\tabularnewline
		& $(2\sigma_u)^2(1\pi_u)^2(3\sigma_g)^1$		& $1^2\Sigma_g^-$	& \tnote{a} 	& \tnote{a}		& 15.15 	& 0.0000	& 15.00	& 0.0000	& 14.51 	& 14.15\tabularnewline
		& $(2\sigma_u)^2(1\pi_u)^2(3\sigma_g)^1$		& $1^2\Sigma_g^+$	& 11.43 		& 0.0004 		& 14.82 	& 0.0009	& 14.67	& 0.0008	& 14.25 	& 14.29\tabularnewline
		& $(2\sigma_u)^1(1\pi_u)^4$					& $1^2\Sigma_u^+$	& 13.95 		& 0.8738 		& 15.33	& 0.7324	& 15.06	& 0.7188	& 15.34 	& 15.09\tabularnewline
		& $(2\sigma_u)^2(1\pi_u)^1(3\sigma_g)^2$		& $2^2\Pi_u$		& \tnote{a} 	& \tnote{a} 	& 14.91	& 0.0172	& 14.71	& 0.0170	& 14.56 	& 15.43\tabularnewline
\hline
$\mae^{prim}$ 	& 									& 				& 0.53 		&  			& 0.28 	& 		& 0.25	& 		& 0.25 	&  \tabularnewline
$\std^{prim}$ 	& 									& 				& 0.71 		&  			& 0.34 	& 		& 0.29	& 		& 0.30 	&  \tabularnewline
$\mae^{sat}$ 	& 									& 				& 2.49 		&  			& 0.66 	& 		& 0.54	& 		& 0.34 	&  \tabularnewline
$\std^{sat}$ 	& 									& 				& 2.68 		&  			& 0.57 	& 		& 0.58	& 		& 0.47 	&  \tabularnewline
\hline	
\hline
\end{tabular}
	\begin{tablenotes}
		\item[a] State is absent in SR-ADC(3).
	\end{tablenotes}
\end{threeparttable}
\end{table*}

\cref{tab:c2} reports accurate EA's and IP's for primary and satellite transitions in the photoelectron spectrum of \ce{C2} computed using SHCI, along with results from SR-ADC(3), MR-ADC(2), MR-ADC(2)-X, and sc-NEVPT2. Out of five EA transitions reported, only one has a single-excitation satellite character [\ce{C2}($X^1\Sigma_g^+$) $\rightarrow$ C$_2^-$ ($1^2\Sigma_u^+$)], while the remaining four correspond to primary peaks. Among six IP's, two have a primary character ($1^2\Pi_u$, $1^2\Sigma_u^+$), three are singly-excited satellites ($1^2\Delta_g$, $1^2\Sigma_g^-$, $1^2\Sigma_g^+$), and one ($2^2\Pi_u$) corresponds to a satellite peak with a double-excitation character. All satellite transitions can be easily identified in the SR-ADC and MR-ADC computations as charged excitations with small spectroscopic factors ($P$ $<$ 0.5). 

For the primary EA's and IP's, the best performance is demonstrated by MR-ADC(2)-X and sc-NEVPT2 with mean absolute errors ($\mae^{prim}$) of 0.25 eV and a standard deviation ($\std^{prim}$) of 0.3 eV, relative to SHCI. The MR-ADC(2) method shows somewhat larger errors ($\mae^{prim}$ = 0.28 eV, $\std^{prim}$ = 0.34 eV), while SR-ADC(3) exhibits the poorest performance ($\mae^{prim}$ = 0.53 eV, $\std^{prim}$ = 0.71 eV) as expected from the single-reference nature of this approximation. Although MR-ADC(2)-X and sc-NEVPT2 perform similarly on average, the former method shows a much better agreement with SHCI for the \ce{C2} band gap computed as energy spacing between the lowest IP and EA peaks in the photoelectron spectrum, with errors of 0.05 and 0.33 eV for MR-ADC(2)-X and sc-NEVPT2, respectively.

All four methods show significantly larger errors for the satellite transitions. The fact that the states involved in these transitions have challenging electronic structures is demonstrated by the performance of SR-ADC(3), which predicts only three out of five satellite transitions and yields very large mean absolute and standard deviation errors ($\mae^{sat}$ and $\std^{sat}$ $>$ 2 eV). The MR-ADC(2) method shows a significant improvement over SR-ADC(3) ($\mae^{sat}$ = 0.66 eV and $\std^{sat}$ = 0.57 eV), while MR-ADC(2)-X outperforms MR-ADC(2) with $\mae^{sat}$ = 0.54 eV and $\std^{sat}$ = 0.58 eV. The best agreement with SHCI for the satellite transitions is demonstrated by sc-NEVPT2 with $\mae^{sat}$ = 0.34 eV and $\std^{sat}$ = 0.47 eV. Both MR-ADC(2)-X and sc-NEVPT2 produce an incorrect order of the $1^2\Sigma_g^-$ and $1^2\Sigma_g^+$ electronic states of C$_2^+$ and large errors in IP for the doubly-excited satellite $2^2\Pi_u$ peak (0.72 and 0.87 eV for MR-ADC(2)-X and sc-NEVPT2, respectively). 

\subsection{Twisted Ethylene}
\label{sec:results:twisted_ethylene}

\begin{table*}[t!]
\caption{Twisted ethylene vertical electron attachment (\tethm) and ionization (\tethp) energies ($\Omega$, eV) and spectroscopic factors ($P$). For MR-ADC and sc-NEVPT2, the CASSCF reference wavefunction was computed using the (8e, 12o) active space. Also shown are mean absolute errors (\mae) and standard deviations (\std) in transition energies, relative to SHCI.}
\label{tab:t-c2h4}
\setstretch{1}
\scriptsize
\centering
%\begin{tabular}{L{1cm}C{1cm}C{1cm}C{1cm}C{0.1cm}C{1cm}C{1cm}C{0.1cm}C{1cm}C{0.1cm}C{0.1cm}C{2cm}C{0.1cm}C{1cm}}
\begin{threeparttable}
\begin{tabular}{lcccccccccc}
\hline
\hline
Molecule & Configuration & State & \multicolumn{2}{c}{SR-ADC(3)} & \multicolumn{2}{c}{MR-ADC(2)} & \multicolumn{2}{c}{MR-ADC(2)-X} & sc-NEVPT2 & SHCI\tabularnewline
  &  & & $\Omega$ & $P$ & $\Omega$ & $P$ & $\Omega$ & $P$ & $\Omega$ & $\Omega$\tabularnewline
\hline
\tethm 	& $(1e)^4(3a_1)^2(2e)^3$			& $1^2E$		& $0.78$	& 0.71	& $-0.40$	& 0.44	& $-0.25$	& 0.44	& $-0.78$	& $-0.43$ \tabularnewline
		& $(1e)^4(3a_1)^2(2e)^2(1a_2)^1$	& $1^2B_2$	& $-1.55$	& 0.85	& $-2.00$	& 0.99	& $-1.94$	& 0.98	& $-1.91$	& $-1.60$ \tabularnewline
		& $(1e)^4(3a_1)^2(2e)^2(4a_1)^1$	& $1^2B_1$	& $-1.83$	& 0.88	& $-2.18$	& 0.63	& $-1.99$	& 0.63	& $-2.03$	& $-1.99$	\tabularnewline		
\hline
\tethp	& $(1e)^4(3a_1)^2(2e)^1$			& $1^2E$	 	& 7.61	& 0.87	& 8.99	& 0.46	& 8.78	& 0.45	& 8.71	& 8.88  \tabularnewline
		& $(1e)^4(3a_1)^1(2e)^2$			& $1^2B_1$	& 14.35	& 0.89	& 14.47	& 0.90	& 14.13 	& 0.88 	& 14.38	& 14.29 \tabularnewline
		& $(1e)^3(3a_1)^2(2e)^2$			& $2^2E$	 	& 14.57	& 0.70	& 15.15	& 0.91	& 14.95	& 0.89	& 14.93	& 14.99 \tabularnewline
\hline
\mae 	& 							& 			& 0.53 	&  		& 0.18 	& 		& 0.14	& 		& 0.17 	&  \tabularnewline
\std	 	& 							& 			& 0.81 	&  		& 0.23 	& 		& 0.17	& 		& 0.17 	&  \tabularnewline
\hline	
\hline
\end{tabular}
%	\begin{tablenotes}
%		\item[a] State is absent in SR-ADC(3).
%	\end{tablenotes}
\end{threeparttable}
\end{table*}

%Brooks & Schaefer - https://pubs.acs.org/doi/pdf/10.1021/ja00496a005
Finally, we investigate performance of MR-ADC(2) and MR-ADC(2)-X for EA and IP of the ethylene molecule at twisted geometry with a 90.0\degree dihedral \ce{H-C-C-H} angle. Twisted ethylene (\teth) is a classic example of a system, which lowest-energy singlet electronic state ($N^1B_1$, originating from the $(1e)^4(3a_1)^2(2e)^2$ valence electronic configuration) has a wavefunction that is dominated by two equally-important Slater determinants [$(2e_x)^2(2e_y)^0$ and $(2e_x)^0(2e_y)^2$].\cite{Merer:1969p639,Brooks:1979p307,BenNun:2000p237,Krylov:2001p375,Barbatti:2004p11614} Since each determinant has a contribution of about 50\% to the wavefunction, the natural occupancies of the \teth frontier orbitals $n(2e_x)$ and $n(2e_y$) $\approx$ 1.

\cref{tab:t-c2h4} compares results of SR-ADC(3), MR-ADC(2), MR-ADC(2)-X, and sc-NEVPT2 with accurate EA's and IP's from SHCI. For the first EA and IP, the best agreement with SHCI is shown by MR-ADC(2) and MR-ADC(2)-X with errors of less than 0.2 and 0.15 eV, respectively. The sc-NEVPT2 method produces a larger error for EA (0.35 eV) and a similar error for IP (0.17 eV). Both MR-ADC(2) and MR-ADC(2)-X correctly describe open-shell nature of the frontier $2e_x$ and $2e_y$ orbitals yielding spectroscopic factors $P$ $\approx$ 0.5, which indicates that the corresponding single-particle states are nearly half-occupied. Importance of multireference effects for the first EA and IP of \teth is demonstrated by the poor performance of SR-ADC(3) that predicts large spectroscopic factors ($P$ $\sim$ 0.7 - 0.9) and overestimates EA and IP from SHCI by 1.21 and 1.27 eV, respectively, incorrectly predicting a bound electronic state for \tethm. Overall, for the three lowest-energy EA's and IP's of \teth, the best results are shown by MR-ADC(2)-X (\mae = 0.14 eV and \std = 0.17 eV), with sc-NEVPT2 (\mae = 0.17 eV and \std = 0.17 eV) and MR-ADC(2) (\mae = 0.18 eV and \std = 0.23 eV) showing somewhat larger errors, on average. 

\section{Conclusions}
\label{sec:conclusions}

In this work, we presented a new implementation and benchmark of multireference algebraic diagrammatic construction theory for electron attachment and ionization (EA/IP-MR-ADC). Following our earlier work on the strict second-order IP-MR-ADC approach (IP-MR-ADC(2)),\cite{Chatterjee:2019p5908} we report the first implementation of the second-order EA-MR-ADC(2) method, as well as the extended EA/IP-MR-ADC(2)-X approximations that partially incorporate third-order correlation effects in the calculation of transition energies and properties. Taking advantage of a small approximation for the second-order amplitudes of the effective Hamiltonian, our implementation of both EA/IP-MR-ADC(2) and EA/IP-MR-ADC(2)-X has the same $\mathcal{O}(M^5)$ computational scaling with the basis set size $M$ and a fixed active space as that of the single-reference EA/IP-ADC(2) method. Additionally, by refactoring terms in the equations and constructing efficient intermediates, our EA/IP-MR-ADC(2)-X implementation completely avoids calculating the four-particle reduced density matrices (4-RDMs), offering a lower $\mathcal{O}(N_{\mathrm{det}} N^6_{\mathrm{act}})$ scaling with the size of the active space ($N_{\mathrm{act}}$) compared to the $\mathcal{O}(N_{\mathrm{det}} N^8_{\mathrm{act}})$ scaling of conventional implementations of second-order multireference perturbation theories.

We benchmarked performance of EA/IP-MR-ADC(2)-X for a set of eight small molecules at equilibrium and stretched geometries, carbon dimer (\ce{C2}), and a twisted ethylene molecule (\teth). To ensure a consistent benchmark, the errors in the EA/IP-MR-ADC(2)-X electron affinities (EA's) and ionization energies (IP's) were calculated relative to accurate EA's and IP's from semi-stochastic heat-bath configuration interaction (SHCI) extrapolated to the full configuration interaction limit. In all tests, the accuracy of the extended EA/IP-MR-ADC(2)-X approximations was found to be similar to that of strongly-contracted N-electron valence second-order perturbation theory (sc-NEVPT2). In particular, MR-ADC(2)-X outperformed sc-NEVPT2 for EA's of small molecules, as well as EA's and IP's of \teth, while sc-NEVPT2 showed smaller errors for IP's of small molecules and satellite transitions of \ce{C2}. 

Importantly, while sc-NEVPT2 requires separate calculations for each electronic state of the neutral and electron-attached/ionized molecules, EA/IP-MR-ADC(2)-X provide a direct access to many EA's and IP's in a single calculation. This, coupled with EA/IP-MR-ADC(2)-X ability to efficiently calculate spectroscopic properties (e.g., spectroscopic factors and density of states), makes EA/IP-MR-ADC(2)-X an attractive alternative to sc-NEVPT2 for calculations of charged excitation energies and photoelectron spectra of multireference systems. To realize the full potential of the EA/IP-MR-ADC(2)-X methods, an efficient implementation of these methods will be developed in our future work. We also plan on extending our current implementation to calculations of open-shell and multireference systems with a large number of active orbitals by combining EA/IP-MR-ADC(2)-X with density matrix renormalization group and selected configuration interaction reference wavefunctions. Finally, developing an implementation of IP-MR-ADC(2)-X for ionizations of core electrons is another avenue that we are planning to explore. 

\section{Appendix: Avoiding 4-RDM in pc-NEVPT2 and MR-ADC(2) Amplitude Equations}
\label{sec:appendix}

Solving the partially-contracted NEVPT2 (pc-NEVPT2)\cite{Angeli:2001p10252,Angeli:2001p297,Angeli:2004p4043} and MR-ADC(2)\cite{Sokolov:2018p204113,Chatterjee:2019p5908} equations for the semi-internal double-excitation amplitudes $t_{ix}^{yz(1)}$ and $t_{xy}^{az(1)}$ requires computation of the zeroth-order Hamiltonian matrix elements of the form
\begin{align}
	\label{eq:K_p1p}
	K_{xyz,uvw}^{[+1']} &= \braket{\Psi_0| a_{yz}^{x} [H^{(0)}, a_{u}^{vw}] |\Psi_0} \\
	\label{eq:K_m1p}
	K_{xyz,uvw}^{[-1']} &= \braket{\Psi_0| a_{z}^{xy} [H^{(0)}, a_{uv}^{w}] |\Psi_0} 
\end{align}
where $a_{r}^{pq} \equiv \c{p}\c{q}\a{r}$ and $a_{qr}^{p} \equiv \c{p}\a{r}\a{q}$. Formally, \cref{eq:K_p1p,eq:K_m1p} depend on the four-particle reduced matrix (4-RDM) of the reference wavefunction $\ket{\Psi_0}$, which has a high $\mathcal{O}(N_{\mathrm{det}} N^8_{\mathrm{act}})$ computational scaling with the number of active orbitals $N_{\mathrm{act}}$ and becomes prohibitively expensive for large active spaces (with $N_{\mathrm{act}}$ $\ge$ 18). A number of techniques have been proposed for evaluating contributions of the $t_{ix}^{yz(1)}$ and $t_{xy}^{az(1)}$ amplitudes without computing 4-RDM in strongly-contracted and uncontracted NEVPT2.\cite{Sharma:2014p111101,Sokolov:2016p064102,Roemelt:2016p204113,Sokolov:2017p244102,Sharma:2017p488,Mahajan:2019p211102} Here, we demonstrate that the matrix elements $K_{xyz,uvw}^{[+1']}$ and $K_{xyz,uvw}^{[-1']}$ (and, thus, the corresponding amplitudes $t_{ix}^{yz(1)}$ and $t_{xy}^{az(1)}$) in pc-NEVPT2 and MR-ADC(2) can be evaluated without computing 4-RDM and introducing any approximations, with a lower scaling. As an example, we consider the terms in \cref{eq:K_p1p} that depend on 4-RDM
\begin{align}
	\label{eq:K_p1p_4rdm_terms}
	%K22 += 0.5 * np.einsum('Vxyz, WXyzUYZx->XYZUVW', v_aaaa_so, rdm_ccccaaaa_so
	K_{xyz,uvw}^{[+1']} \Leftarrow &\frac{1}{2} \sum_{x'y'z'} \v{vx'}{y'z'} \braket{\Psi_0| \c{w} \c{x} \c{y'} \c{z'} \a{u} \a{y} \a{z} \a{x'} |\Psi_0} \notag \\
	%K22 -= 0.5 * np.einsum('Wxyz, VXyzUYZx->XYZUVW', v_aaaa_so, rdm_ccccaaaa_so
	 - &\frac{1}{2} \sum_{x'y'z'} \v{wx'}{y'z'} \braket{\Psi_0| \c{v} \c{x} \c{y'} \c{z'} \a{u} \a{y} \a{z} \a{x'} |\Psi_0} \notag \\
	%K22 -= 0.5 * np.einsum('xyUz, VWXzYZxy->XYZUVW', v_aaaa_so, rdm_ccccaaaa_so	
	 - &\frac{1}{2} \sum_{x'y'z'} \v{x'y'}{uz'} \braket{\Psi_0| \c{v} \c{w} \c{x} \c{z'} \a{y} \a{z} \a{x'} \a{y'} |\Psi_0} 
\end{align}
where we omitted contributions of lower-rank RDMs for clarity. Reordering creation and annihilation operators such that the operators with summation labels go first,\cite{Sokolov:2016p064102,Sokolov:2017p244102,Sokolov:2018p204113} the first term of \cref{eq:K_p1p_4rdm_terms} can be rewritten as
\begin{align}
	\label{eq:K_p1p_4rdm_term_1}
	&\frac{1}{2} \sum_{x'y'z'} \v{vx'}{y'z'} \braket{\Psi_0| \c{w} \c{x} \c{y'} \c{z'} \a{u} \a{y} \a{z} \a{x'} |\Psi_0} \notag \\
%	&= - \frac{1}{2} \sum_{x'y'z'} \v{vx'}{y'z'} \braket{\Psi_0| \c{y'} \c{z'} \a{x'} \c{w} \c{x} \a{u} \a{y} \a{z} |\Psi_0} \notag \\
%	&+\frac{1}{2} \sum_{y'z'} \v{vw}{y'z'} \braket{\Psi_0| \c{y'} \c{z'} \c{x} \a{u} \a{y} \a{z} |\Psi_0} \notag \\
%	&-\frac{1}{2} \sum_{y'z'} \v{vx}{y'z'} \braket{\Psi_0| \c{y'} \c{z'} \c{w} \a{u} \a{y} \a{z} |\Psi_0} \notag \\
	&= - \frac{1}{2} \braket{v^v| \c{w} \c{x} \a{u} \a{y} \a{z} |\Psi_0} \notag \\
	&+\frac{1}{2} \sum_{y'z'} \v{vw}{y'z'} \braket{\Psi_0| \c{y'} \c{z'} \c{x} \a{u} \a{y} \a{z} |\Psi_0} \notag \\
	&-\frac{1}{2} \sum_{y'z'} \v{vx}{y'z'} \braket{\Psi_0| \c{y'} \c{z'} \c{w} \a{u} \a{y} \a{z} |\Psi_0} 
\end{align}
where we defined intermediate states
\begin{align}
	\label{eq:K_p1p_int_state}
	&\ket{v^v} = \sum_{x'y'z'} \v{y'z'}{vx'} \c{x'} \a{z'} \a{y'} \ket{\Psi_0} 
\end{align}
\cref{eq:K_p1p_4rdm_term_1} demonstrates that the first 4-RDM contribution to \cref{eq:K_p1p_4rdm_terms} can be evaluated without explicitly computing and storing 4-RDM. By precomputing the intermediate states $\ket{v^v}$ with $\mathcal{O}(N_{\mathrm{det}} N^4_{\mathrm{act}})$ active-space scaling, the cost of computing the 4-RDM contribution in \cref{eq:K_p1p_4rdm_term_1} is lowered from $\mathcal{O}(N_{\mathrm{det}} N^8_{\mathrm{act}})$ to $\mathcal{O}(N_{\mathrm{det}} N^6_{\mathrm{act}})$, which is equivalent to the cost of computing 3-RDM. (Note that the last two terms in \cref{eq:K_p1p_4rdm_term_1} have $\mathcal{O}(N^8_{\mathrm{act}})$ scaling, which is significantly lower than $\mathcal{O}(N_{\mathrm{det}} N^6_{\mathrm{act}})$). The remaining 4-RDM contributions in $K_{xyz,uvw}^{[+1']} $ and $K_{xyz,uvw}^{[-1']}$ can be evaluated in a similar way without computing and storing 4-RDM. Importantly, all of these contributions can be expressed in terms of a single set of intermediate states $\ket{v^v}$ defined in \cref{eq:K_p1p_int_state}, which can be computed at the beginning of the calculation and reused for efficient evaluation of $K_{xyz,uvw}^{[+1']} $ and $K_{xyz,uvw}^{[-1']}$. We note that the factorization described here has also been used in the efficient implementation of internally-contracted multireference configuration interaction theory.\cite{Werner:1987p1,Werner:1988p5803}

\acknowledgement
This work was supported by start-up funds provided by the Ohio State University. Computations were performed at the Ohio Supercomputer Center under the project PAS1583.\cite{OhioSupercomputerCenter1987} 

\suppinfo
Comparison of MR-ADC(2) and MR-ADC(2)-X results with exact and approximate second-order amplitudes. Equations of the EA/IP-MR-ADC(2) and EA/IP-MR-ADC(2)-X methods for the $\mathbf{M}$, $\mathbf{T}$, and $\mathbf{S}$ matrix elements. Optimized geometry of the twisted ethylene molecule.

%\bibliography{papers,refs}

\providecommand{\latin}[1]{#1}
\makeatletter
\providecommand{\doi}
  {\begingroup\let\do\@makeother\dospecials
  \catcode`\{=1 \catcode`\}=2 \doi@aux}
\providecommand{\doi@aux}[1]{\endgroup\texttt{#1}}
\makeatother
\providecommand*\mcitethebibliography{\thebibliography}
\csname @ifundefined\endcsname{endmcitethebibliography}
  {\let\endmcitethebibliography\endthebibliography}{}

\end{document}